\documentclass[]{aa}
\usepackage{graphics}
\usepackage{epsf}

\begin{document}

\thesaurus{04(11.04.1; 08.22.1)}

\title{ The extra-galactic Cepheid distance scale from LMC \\
and Galactic period-luminosity relations }

\author{S. M. Kanbur\inst{1} 
\and C. Ngeow\inst{1}
\and S. Nikolaev\inst{2}
\and N. R. Tanvir\inst{3}
\and M. A. Hendry\inst{4} }

\offprints{S. Kanbur \\
 email: {\tt shashi@astro.umass.edu}}

\institute{Astronomy Department, University of Massachusetts, \\
Amherst, MA  01003, USA 
\and Institute for Geophysics and Planetary Physics, \\
Lawrence Livermore National Laboratory, Livermore, CA 94550, USA
\and Department of Physical Science, University of Hertfordshire, \\ 
College Lane, Hatfield, AL10 9AB, UK
\and Department of Physics and Astronomy, University of Glasgow, \\
Glasgow G12 8QQ, UK}

\date{Received 07 May 2003 / Accepted 14 Aug 2003}

\titlerunning{Cepheid distance from LMC and Galactic PL relations}
\authorrunning{Kanbur et al.}
\maketitle

\begin{abstract}

	In this paper, we recalibrate the Cepheid distance to some nearby galaxies observed by the {\it HST} Key Project and the Sandage-Tammann-Saha group. We use much of the Key Project methodology in our analysis but apply new techniques, based on Fourier methods to estimate the mean of a sparsely sampled Cepheid light curve, to published extra-galactic Cepheid data. We also apply different calibrating PL relations to estimate Cepheid distances, and investigate the sensitivity of the distance moduli to the adopted calibrating PL relation. We re-determine the OGLE LMC PL relations using a more conservative approach and also study the effect of using Galactic PL relations on the distance scale.

	For the Key Project galaxies after accounting for charge transfer effects, we find good agreement with an average discrepancy of -0.002 and 0.075$mag.$ when using the LMC and Galaxy, respectively, as a calibrating PL relation. For NGC 4258 which has a geometric distance of $29.28mag.$, we find a distance modulus of $29.44\pm0.06(\mathrm{random})mag.$, after correcting for metallicity. In addition we have calculated the Cepheid distance to 8 galaxies observed by the Sandage-Tammann-Saha group and find shorter distance moduli by $-0.178mag.$ (mainly due to the use of different LMC PL relations) and $-0.108mag.$ on average again when using the LMC and Galaxy, respectively, as a calibrating PL relation. However care must be taken to extrapolate these changed distances to changes in the resulting values of the Hubble constant because STS also use distances to NGC 3368 and 4414 and because STS calibration of SN Ia is often decoupled from the distance to the host galaxy through their use of differential extinction arguments. We also calculate the distance to all these galaxies using PL relations at maximum light and find very good agreement with mean light PL distances.

	However, after correcting for metallicity effects, the difference between the distance moduli obtained using the two sets of calibrating PL relations becomes negligible. This suggests that Cepheids in the LMC and Galaxy do follow different PL relations and constrains the sign for the coefficient of the metallicity correction, $\gamma$, to be negative, at least at the median period $\log(P) \approx 1.4$, of the target galaxies. 


\end{abstract}

%***********************************************************
%   SECTION 1: INTRODUCTION
%***********************************************************

\section{Introduction}

	The extra-galactic distance scale is one of the key problems in modern astronomy. One of the basic parts of the solution is the correlation between period and mean luminosity (PL) obeyed by classical Cepheids. The Hubble Space Telescope ({\it HST}) $H_0$ Key Project (Freedman et al. 2001, hereafter KP) has used Cepheids (discovered either by ground-based telescopes or by {\it HST}) in 31 spiral galaxies, with 18 of them observed originally by KP, and the PL relation to estimate the distances to these galaxies. These distances were then used in turn to calibrate a host of secondary distance indicators and hence estimate Hubble's constant. In parallel with this a team led by Sandage et al. (Saha et al. 2001b, hereafter STS) has used the {\it HST} to discover Cepheids in spiral galaxies which host Type Ia supernovae. Cepheid distances to these galaxies were used to calibrate the Hubble diagram for Type Ia supernovae and hence estimate Hubble's constant.

	Though both groups used mostly {\it HST} observations to discover the Cepheids and similar methodologies, the KP team favor a short distance scale and a larger value of $H_0$ ($72 \pm 8 kms^{-1}Mpc^{-1}$, Freedman et al. \cite{kp}) whilst the STS group favor a long distance scale and smaller value of $H_0$ ($58.7 \pm 6.3 kms^{-1}Mpc^{-1}$, Saha et al. \cite{sah01b}). The discrepancy in the value of the Hubble constant from these two groups is still unresolved (e.g., see Hendry \& Kanbur \cite{hen96}; Beaulieu et al. \cite{bea97}; Kochanek \cite{koc97}; Sasselov et al. \cite{sas97}; Tanvir et al. \cite{tan99b}; Caputo et al. \cite{cap00}; Gibson et al. \cite{gib00}; Gibson \& Stetson \cite{gib01}). It is due only in part to the Cepheid distance scale. We note that the recent WMAP results (Spergel et al. \cite{spe03}) find a Hubble constant very similar to the value obtained by KP. Nevertheless it is still very instructive to discover the reason behind the discrepancies in the Cepheid distance scale from the STS and KP groups. A proper understanding of the Cepheid PL relation in this regard is very important for an accurate local distance scale. In this paper we concentrate on recalibrating the Cepheid distance to the target galaxies in both KP and STS groups with existing data because of the following factors:
	\begin{enumerate}
	\item 	{\it HST} Cepheid data is sparsely sampled with typically 12 and 4 points per Cepheid in the V and I band respectively (see Section 2 for details). In order to apply the Cepheid PL relations, it is necessary to estimate the mean magnitudes from these data in both V and I bands. In this paper, we use the techniques of Fourier expansion and interrelations to estimate the V and I mean magnitudes from sparsely populated light curves. These techniques are described in detail in Ngeow et al. (\cite{nge03}).
	\item 	We re-analyze the OGLE (Optical Gravitational Lensing Experiment) LMC Cepheids (Udalski et al. \cite{uda99a}) on the basis of the quality of their $V$ band light curve, and develop more conservative LMC PL relations based on this analysis and investigate the sensitivity of extra-galactic Cepheid distances to the LMC PL relation. This approach is different to the ``sigma-clipping'' methods currently used by Udalski et al. (\cite{uda99a}). 
	\item	The average value of  metallicities, defined as $12+\log(O/H)$, in target galaxies is about $8.84\pm0.31dex$ (Freedman et al. \cite{kp}). This value is closer to the standard Solar value of $8.87\pm0.07dex$ (Grevesse et al. \cite{gre96}) than the LMC value of $8.50\pm0.08dex$ (see reference in Ferrarese et al. \cite{fre00}). Therefore, another approach is to use Galactic Cepheid PL relations as fundamental calibrating relations (Feast \cite{fea01,fea03}; Tammann \cite{tam03a}; Thim et al. \cite{thim03}; Fouqu\'{e} et al. \cite{fsg}). We compare the distances obtained when using both the LMC and Galactic PL relations. 
	\item	Kanbur \& Hendry (\cite{kan96}) building on the hydrodynamical models of Simon et al. (\cite{sim93}) suggested that PL relations at maximum light may be more accurate than PL(mean) relations. To test this idea, we compute distances using PL(max) relations and compare with their mean light counterparts.
	\end{enumerate}

	In Section 2 we summarize the photometric data used in this study. In Section 3 we present and discuss the methodology used in determining Cepheid distances, including obtaining the V and I band means and applying different PL relations. The results will be presented in Section 4. The conclusion and the discussion will be presented in Section 5.

%******************************************************
%   SECTION 2: PHOTOMETRIC DATA
%*****************************************************

\section{The Photometric Data}

	The photometry data for the Cepheids in each target galaxy were directly obtained from the corresponding published papers. We emphasize that we did not repeat the photometric reduction from raw data. The target galaxies that are selected in this study include: 16 KP galaxies, 8 STS galaxies and NGC 4258, which has an accurate geometrical distance measurement from water maser studies (Herrnstein et al. \cite{her99}, hereafter WM galaxy). The photometric data of 16 KP galaxies were directly downloaded from the KP web-page\footnote{The URL is http://www.ipac.caltech.edu/H0kp/H0KeyProj .html. The reference to each KP galaxies also listed in this URL.}, excluding NGC 1425 (the data is not available at the time of analysis) and NGC 5457 (M101, as the observations to this galaxy include its outer field (Kelson et al. \cite{kel96}) and inner field (Stetson et al. \cite{ste98b}), which complicated the analysis). The photometric data for the STS and WM galaxies were taken from the STS papers (see Saha et al. \cite{sah01b} for references for each target galaxy)\footnote{We exclude NGC 3368 (Tanvir et al. \cite{tan99b}) because it's not part of the STS program, and we are not calibrating the SN Ia distance in this paper.} and Newman et al. (\cite{new01}), respectively. The list of the target galaxies can be obtained from Table \ref{tab2}.

	There are two major photometry reduction packages used in reducing the data for these galaxies, the ALLFRAME (Stetson \cite{ste94,ste96}) and a variant of the DoPHOT (Schechter et al. \cite{sch93}; Saha et al. \cite{sah96}) package. The KP team utilized both packages as a double-blind reduction process (Kennicutt et al. \cite{ken95}; Freedman et al. \cite{kp}) to check the consistency between the two packages. However, the final, adopted distance moduli are based on ALLFRAME results (Freedman et al. \cite{kp}). Inter-comparison of the results from these two packages (galaxy-by-galaxy basis) show good agreement between the two (see the KP papers for more details and Hill et al. \cite{hil98}). We also use the ALLFRAME photometric data downloaded from their web-page (if available). However, the photometric data for NGC 2541 and NGC 4321 are only available in DoPHOT from the same web-page. Similarly, only the DoPHOT photometric data is available for WM galaxy, although the photometry reductions were carried out by both packages (Newman et al. \cite{new01}). 
	 
	In contrast, the photometric data for the STS galaxies were reduced mainly with DoPHOT, although some of the galaxies also used an additional reduction package to check the DoPHOT results (e.g., reduction of NGC 4527 included the ROMAFOT package, see Saha et al. \cite{sah01a}). Nevertheless, the available data for all STS galaxies are from the DoPHOT package. Although the KP team has reanalyzed the STS galaxies with the ALLFRAME package (Gibson et al. \cite{gib00}; Gibson \& Stetson \cite{gib01}; Stetson \& Gibson \cite{ste01}; Freedman et al. \cite{kp}), these data are not yet available from the KP web-page (as claimed in these papers). However, most, but not all, of the Cepheids are common in both STS and KP results (Gibson et al. \cite{gib00}). Hence, we only can use the DoPHOT results in our analysis of STS galaxies. 

	The next steps in the photometric reduction process are a Cepheid search and period determination. Each reduction package has its own algorithms to perform the period search, and the results generally agree well. We use published Cepheid data and periods in our analysis.

	The {\it HST} observations of the target galaxies generally consist of 24 V band and 8 I band cosmic-ray (CR) split images. The analysis of DoPHOT and ALLFRAME treat these images differently to deal with the cosmic-ray and point-spread functions. However, most of the published photometry combined the back-to-back CR split images together and tabulated the averaged 12 V band and 4 I band data points. Therefore, we only can use these (reduced) photometric data points to reconstruct the light curves and obtain the mean magnitudes by the Fourier techniques described in the next section. 

%******************************************************
%   SECTION 3: GENERAL METHODOLOGY AND ANALYSIS
%*****************************************************

\section{General Methodology and Analysis}

	The physical basis of Cepheid PL relations and their usage in determining the distance have been covered extensively in the literature (e.g., see Sandage \& Tammann \cite{san68}; Feast \& Walker \cite{fea87}, Madore \& Freedman \cite{mad91}; Freedman et al. \cite{kp}) and would not be repeated here, only a brief description will be presented. The application of the Cepheid PL relation to estimate distances involves the discovery and appropriate observation of Cepheids sufficient to estimate their periods (in days) and mean magnitudes. In order to correct for the extinction/reddening, observations of extra-galactic Cepheids by {\it HST} are normally taken in bandpasses V and I. The PL relation in bandpass $\lambda$ can be expressed as: $M_\lambda=a_\lambda\log(P)+b_\lambda$, where $a$ and $b$ are the slope and zero-point, respectively. For the measurement of the apparent mean magnitudes of Cepheids in a target galaxy, $<m_\lambda>$, the distance modulus in bandpass $\lambda$ is: 

	\[ \mu_\lambda = <m_\lambda> - a_\lambda\log(P) - b_\lambda. \] 

	Since $\mu_0=\mu_V-A_V=\mu_I-A_I$, the reddening-free distance modulus, or the Wessenheit function (Madore \cite{mad82}; Moffett \& Barnes \cite{mof86}; Madore \& Freedman \cite{mad91}; Tanvir \cite{tan99}), can be derived as:

	\begin{eqnarray}
	\mu_{0} & = & \mu_V - R (\mu_V - \mu_I), 
	\end{eqnarray}

	where $R\equiv A_V/E(V-I)$ is the ratio of total-to-selective absorption. Following KP, we adopt $R=2.45$ from Cardelli et al. (\cite{car89}). Note that the validity of equation (1) is based on the assumption that there are no correlated measurement errors in the V and I bandpasses. Then any differences in the V and I band distance moduli are due to differential reddening (see Saha et al. \cite{sah96,sah01b} for details).

	After applying a period cut to the short period Cepheids in target galaxies (to avoid the incompleteness bias at the faint-end of observations, see Sandage \cite{san88}; Lanoix et al. \cite{lan99}; Freedman et al. \cite{kp}), an unweighted mean of apparent distance moduli to the remaining Cepheids is adopted as the distance modulus to the target galaxy. The random (or statistical) error associated with equation (1) can be calculated via the standard formulae (i.e. $\sigma/\sqrt{N}$). The reasons for taking the unweighted mean are: (a) the photometric errors of the mean magnitudes are smaller than the expected width of the instability strip, and are, together with other systematic errors, the dominant part of the weight in a weighted mean (Leonard et al. \cite{leo03}); (b) this is equivalent to the fitting scheme used by KP; and (c) it can be easily incorporated with the weighting scheme adopted by the STS group (Tanvir \cite{tan97}). Finally, due to the possible metallicity dependence of the Cepheid PL relation (see the discussions and reference in, e.g., Kennicutt et al. \cite{ken98}; Freedman et al. \cite{kp}), metallicity corrections are added to equation (1):

	\begin{eqnarray}
	\mu_z & = & \mu_0 + \delta_z,
	\end{eqnarray}

	where $\delta_z=\gamma([O/H]_{ref}-[O/H]_{gal})$ with the usual definition of $[O/H]\equiv 12+\log(O/H)$. The reference metallicity, $[O/H]_{ref}$, is $8.50dex$ and $8.87dex$ for using LMC and Galactic PL relations, respectively. Some published values of $\gamma$ range from $-0.88\pm0.16$ and $-0.56\pm0.20$ (Gould \cite{gou94}), $-0.44^{+0.10}_{-0.20}$ (Beaulieu et al. \cite{bea97}; Sasselov et al. \cite{sas97}), $-0.4\pm0.2$ (Kochanek \cite{koc97}), $-0.32\pm0.21$ (Freedman \& Madore \cite{fre90}), $-0.24\pm0.16$ (Kennicutt et al. \cite{ken98}) to $0.27$ (Caputo et al. \cite{cap00}). Note that the value of $\gamma$ is method-dependent, i.e. it depends on the bandpasses and calibrating PL relation adopted. For example, strictly speaking the metallicity dependence quoted in Kennicutt et al. (\cite{ken98}) is only applicable when the Madore \& Freedman (\cite{mad91}) PL relation is used, and then only for the V and I bandpasses. Although there is some debate on the value for $\gamma$ (see Freedman et al. \cite{kp}; Tammann et al. \cite{tam01} for details), we adopt $\gamma \equiv \gamma_{V,I}=-0.2\pm0.2mag\ dex^{-1}$ as in KP, who note that this value is in the middle of a number of different determinations of $\gamma$. Changes in metallicity affect the mean brightness of a Cepheid and the papers quoted above were aimed at quantifying this brightness shift due to metallicity. However, we now know that the Cepheid PL relation has a different slope in the Galaxy and LMC, and indeed may have two different slopes in the LMC itself. Thus it may be that metallicity affects not only the mean brightness of a Cepheid but also the way that mean brightness changes with period (the slope of the PL relation). We discuss this and its implications for metallicity corrections like equation (2) in Section 5.

	Since we are using the published photometry data to calculate the distance modulus, we share the same systematic errors as in the published results. These include the uncertainty in the zero point of the PL relation, calibration of photometric zero points, reddening and metallicity effects, incompleteness bias and crowding. The order of a typical systematic error is around $\sim 0.1-0.2mag$.

\subsection{The Mean Magnitudes}

	As mentioned in Section 2, most of the {\it HST} observations of extra-galactic Cepheids contain only 12 V band epochs and 4 (or 5) I band epochs. These observations use a power-law time series sampling strategy to minimize the aliasing problem and maximize the phase coverage of the observed data points within the observing windows (Freedman et al. \cite{fre94}; Kennicutt et al. \cite{ken95}). In order to reduce the bias due to sampling procedures, STS and some early KP observations used phase-weighted intensity means to find mean magnitudes in the V band. The I band mean magnitudes were found via empirical relations developed by Freedman (\cite{fre88}) and Labhardt et al. (\cite{lab97}) for KP and STS observations, respectively (see, e.g., Silbermann et al. \cite{sil96}). Other KP observations used a template-fitting procedure (Stetson \cite{ste96}) to obtain the mean magnitudes in both V and I band simultaneously.

	In addition to the methods mentioned, Ngeow et al. (\cite{nge03}) recently developed an alternative method to obtain the mean magnitudes, which is based on Fourier techniques. Ngeow et al. (\cite{nge03}) give specific examples of situations when such a method could be preferable to template based techniques. Another motivation for developing these Fourier techniques is the possibility of reconstructing observed HST light curves to compare with models. In Section 4.1, we quantitatively compare the means obtained using our methods with existing techniques in the literature.

	Here, we apply the new Fourier method to all target galaxies in order to be consistent in doing the analysis, and also to test this new method in the distance scale problem. The details of using the Fourier techniques are presented in Ngeow et al. (\cite{nge03}). Here we outline the steps:

	\begin{enumerate}
	\item 	{\it The V band}: The light curves of 12 V band data points are reconstructed via a $4^{th}$-order Fourier expansion with the form:
	\begin{eqnarray}
	m(t) & = & m_0 + \sum^{i=4}_{i=1} [A_i\cos(2i\pi t/P + \phi_i)] 
	\end{eqnarray}

	where $A$ and $\phi$ are Fourier amplitudes and phases, respectively, and $m_0$ is the mean magnitude. Since the periods are directly taken from literature, the remaining nine parameters can be obtained by fitting the data with equation (3). We use the technique of simulated annealing and restrict the ranges that the Fourier parameters in equation (3) can take to reconstruct the V band light curves. The ranges of Fourier phases are from $0$ to $2\pi$, whilst the ranges of Fourier amplitudes are determined from the ``calibrating set'' Cepheids and OGLE LMC Cepheids. The ``calibrating set'' Cepheids consist of $\sim$100 Galactic Cepheids, mostly observed originally by Moffett \& Barnes (\cite{mof84}), and some LMC/SMC Cepheids.   
	\item 	{\it The I band}: For 4 I band data points, it is clear that equation (3) cannot be applied to reconstruct the I band light curves. Therefore, we use  Fourier interrelations (Ngeow et al. \cite{nge03}) to reconstruct the I band light curves. The Fourier interrelations are linear relations connecting the Fourier parameters in the V and I bands:

	\begin{eqnarray}
	A_i(I) & = & \alpha_i +\beta_iA_i(V);\  \ \phi_i(I)=\gamma_i+\eta_i\phi_i(V)
	\end{eqnarray}

	The coefficients in equation (4) are determined from the ``calibrating set'' Cepheids. Ngeow et al. (\cite{nge03}) developed such Fourier interrelations for the Galaxy, LMC and SMC separately and found that they are only weakly dependent on metallicity. This is important since it means that the interrelations can be applied over a wide range of metallicity. Given the solution of equation (4), the observed I band points are then used to establish the I band light curves and hence estimate the corresponding mean magnitudes by minimizing the chi-square.
	\item	These Fourier techniques have been tested with Monte Carlo simulations, and show that the reconstruction procedures are unbiased and the errors of the Fourier amplitudes and means are around $\sim0.03mag$. 
	\end{enumerate}

	Although the filters installed on the {\it HST} are close to the standard Johnson and Kron-Cousins system, i.e. $F555W\sim V$ and $F814W\sim I$, conversions from the {\it HST} passbands to the standard photometry systems have been used by both KP and STS teams. For the case of the ALLFRAME photometry reduction in the KP galaxies, the conversions are made during the reductions with the formulae suggested by Holtzman et al. (\cite{hol95}), hence the published photometry of the Cepheids are in standard bands. A similar situation regarding the conversion holds for the two KP galaxies with DoPHOT photometry (NGC 2541 \& NGC 4321) and the WM galaxy. For STS galaxies with DoPHOT photometric reductions, the conversions are applied after the reductions to the calculated mean magnitudes (see equations (5) and (6) in Saha et al. \cite{sah94} for WFPC; and equations (2) and (3) in Saha et al. \cite{sah96} for WFPC2). In the case of PL(max) relations, however, no such conversion is available yet. We have to assume the conversions at maximum light are similar to the equations given by Saha et al. (\cite{sah94,sah96}), and apply the conversion accordingly. These conversions have been applied to STS galaxies in Table \ref{tab3a}, \ref{tab3b} \& \ref{tab3c}.

	The derived mean magnitudes in both V and I bands for the Cepheids used in this study are listed in Table \ref{taballmag}, where column 1 shows the Cepheid in the target galaxy, column 2 is the period adopted from the literature, and columns 3 and 4 are the mean V and I band magnitudes, respectively, derived from the Fourier techniques described in this paper. Table \ref{taballmag} is available in its complete electronic form at the CDS\footnote{http://cdsweb.u-strasbg.fr/cgi-bin/qcat?J/A+A/}. Here we only show a portion of the table to indicate its form and content.

%*********************************************************************
%  TABLE 1 HERE
%*********************************************************************

	\begin{table}[]
	%\centering
	\caption{Mean V and I magnitudes from Fourier fits.$^{\mathrm a}$}
	\label{taballmag}
	\begin{tabular}{lccc} \hline
	Galaxy-Cepheid  & $\log(P)$ & V(mag.) & I(mag.) \\ 
	\hline \hline
	NGC925-C5 	& 1.686 & 23.680 & 22.478 \\ 
	NGC925-C6 	& 1.635 & 24.533 & 23.453 \\ 
	NGC925-C7 	& 1.624 & 24.305 & 23.457 \\ 
	\hline
	\end{tabular}
	\begin{list}{}{}
	\item $^{\mathrm a}$ The entire table is available electronically at the CDS. 
	\end{list}
	\end{table}

\subsection{The Period-Luminosity Relations}

	In order to apply the PL relations to {\it HST} data, we only look for the published PL relations that are available in both V and I bands in both the Galaxy and LMC. The following subsections discuss the adopted PL relations in this study. Note that for the LMC PL relations, we adopt $\mu_{LMC}=18.50mag.$, to be consistent with the KP team (Freedman et al. \cite{kp}). All of the adopted PL relations are listed in Table \ref{tab1a} and \ref{tab1b}, for the V and I bands, respectively. 

%***************************************************************
%   TABLE 2 & 3 HERE
%***************************************************************

	\begin{table*}[t]
	%\centering
	\caption{The adopted V band Period-Luminosity relations.$^{\mathrm a}$}
	\label{tab1a}
	\begin{tabular}{lcccc} \hline
	Relation & Slope ($a_V$) & ZP ($b_V$) & $\sigma_V$ & $N$ \\ \hline \hline
	LMC (MF91)             & $-2.76\pm0.11$   & $-1.40\pm0.05$   & $0.27$  & 32  \\
	OGLE LMC (U99)         & $-2.760\pm0.031$ & $-1.458\pm0.021$ & $0.159$ & 649 \\
	OGLE LMC (Here)        & $-2.746\pm0.043$ & $-1.401\pm0.030$ & $0.223$ & 634 \\
	LMC PL$_{>10d}$ (TR02) & $-2.48\pm0.17$   & $-1.75\pm0.20$   & $0.16$  & $\sim$47 \\
	LMC PL(Max)            & $-2.744\pm0.051$ & $-1.817\pm0.035$ & $0.261$ & 634 \\ \hline
	Galactic (GFG98)       & $-3.037\pm0.138$ & $-1.021\pm0.040$ & $0.209$ & 28  \\
	Galactic (FSG03)       & $-3.06\pm0.11$   & $-0.989\pm0.034$ & $\cdots$& 32  \\ 
	Galactic (T03)         & $-3.141\pm0.100$ & $-0.826\pm0.119$ & 0.24    & 53  \\ \hline
	\end{tabular} 
	\begin{list}{}{}
	\item	$^{\mathrm a}$ For LMC PL relations, Assume $\mu_{LMC}=18.50 mag.$
	\end{list}
	\end{table*}

	\begin{table*}[t]
	%\centering
	\caption{The adopted I band Period-Luminosity relations.$^{\mathrm a}$}
	\label{tab1b}
	\begin{tabular}{lcccc} \hline
	Relation & Slope ($a_I$) & ZP ($b_I$) & $\sigma_I$ & $N$ \\ \hline \hline
	LMC (MF91)             & $-3.06\pm0.07$   & $-1.81\pm0.03$   & $0.18$  & 32  \\
	OGLE LMC (U99)         & $-2.962\pm0.021$ & $-1.942\pm0.014$ & $0.109$ & 658 \\
	OGLE LMC (Here)        & $-2.965\pm0.028$ & $-1.889\pm0.019$ & $0.145$ & 634 \\ 
	LMC PL$_{>10d}$ (TR02) & $-2.82\pm0.13$   & $-2.09\pm0.15$   & $0.12$  & $\sim$47 \\
	LMC PL(Max)            & $-2.958\pm0.033$ & $-2.129\pm0.023$ & $0.171$ & 634 \\ \hline
	Galactic (GFG98)       & $-3.329\pm0.132$ & $-1.435\pm0.037$ & $0.194$ & 27  \\
	Galactic (FSG03)       & $-3.24\pm0.11$   & $-1.550\pm0.034$ & $\cdots$& 32  \\
	Galactic (T03)         & $-3.408\pm0.095$ & $-1.325\pm0.114$ & 0.23    & 53  \\ \hline
	\end{tabular} 
	\begin{list}{}{}
	\item	$^{\mathrm a}$ For LMC PL relations, Assume $\mu_{LMC}=18.50 mag.$
	\end{list}
	\end{table*}

\subsubsection{The LMC PL Relations}

	The LMC PL relations used by both KP team (before the publication of their final paper) and STS team are based on a homogeneous sample of 32 Cepheids, with periods ranging from 10 days to $\sim$120 days (Madore \& Freedman \cite{mad91}, hereafter MF91). Since this PL relation has been extensively used in determining Cepheid distances in the past, and the STS team still use this PL relation in their study (e.g., see Saha et al. \cite{sah01b}), we adopt the MF91 relations as one of our calibrating PL relations in this study. 

	After the publication of PL relations derived from OGLE LMC Cepheids (Udalski et al. \cite{uda99a}, hereafter U99), the KP team recalibrated their Cepheid distances with these new LMC PL relations in their final paper, as well as the Cepheid distance to NGC 4258 by Newman et al. (\cite{new01}). These reddening-corrected PL relations were derived using Cepheids with $\log P > 0.4$ to minimize possible contamination by first overtone pulsators and were "sigma clipped", resulting in $\sim650$ Cepheids used in deriving the PL relations. In order to compare our results with published Cepheid distances, we also adopt the U99 PL relations. The new U99 PL relations have dramatically changed Cepheid distances compared to distances derived from MF91 PL relations, in the sense that the derived distances are smaller with U99 PL relations (Freedman et al. \cite{kp}).

	There are some criticisms about the U99 PL relations in the literature. The U99 PL relations are dominated by short period Cepheids (with $<\log(P)>\sim 0.5$ and about 90\% of them have period shorter than 10 days), and a lack of longer period Cepheids with $\log(P)>1.5$ (Feast \cite{fea01,fea03}; Saha et al. \cite{sah01b}). However, Freedman et al. (\cite{kp}), by using Sebo et al. (\cite{sebo}) LMC data, claimed that the use of the OGLE LMC PL relations in estimating distances to target galaxies whose Cepheids all had periods longer than the longest period Cepheid observed by Udalski et al. (\cite{uda99a}) made little difference. Another potential problem of the U99 PL relations is the discovery of a break in the PL relation at $\log(P)=1.0$, as shown in Tammann et al. (\cite{tam01}) and in Tammann \& Reindl (\cite{tam02}). The long and short period Cepheids follow different PL relations. The reasons for the break in PL relations are still unclear, nevertheless we adopt the PL relations for long period Cepheids (Tammann \& Reindl \cite{tam02}, hereafter TR02) since most of the extra-galactic Cepheids have periods longer than 10 days.

	Despite these potential problems, we re-analyze OGLE LMC Cepheid data and re-fit the PL relations without using sigma-clipping. The sigma-clipping method is an iterative procedure whereby the data are fitted with a regression and then those points lying 2.5$\sigma$ away from the regression line are removed. A new regression is fitted and the procedure continues for few cycles. Therefore, it blindly removes all the outliers to reduce the scatter in the final fit. However, we feel this approach may both remove some true Cepheids from the data and include some suspicious Cepheids in the final sample (see discussion below). In addition, Nikolaev et al. (\cite{nik03}) discussed some flaws associated with the sigma-clipping algorithm, including the implicit assumption of the normal distribution of the residuals and the sensitivity of the results to the chosen threshold of $k\sigma$.

%**********************************************************
%      FIGURE 1 HERE
%**********************************************************

	\begin{figure*}
	\vspace{0cm}
	\hbox{\hspace{0.2cm}\epsfxsize=7.5cm \epsfbox{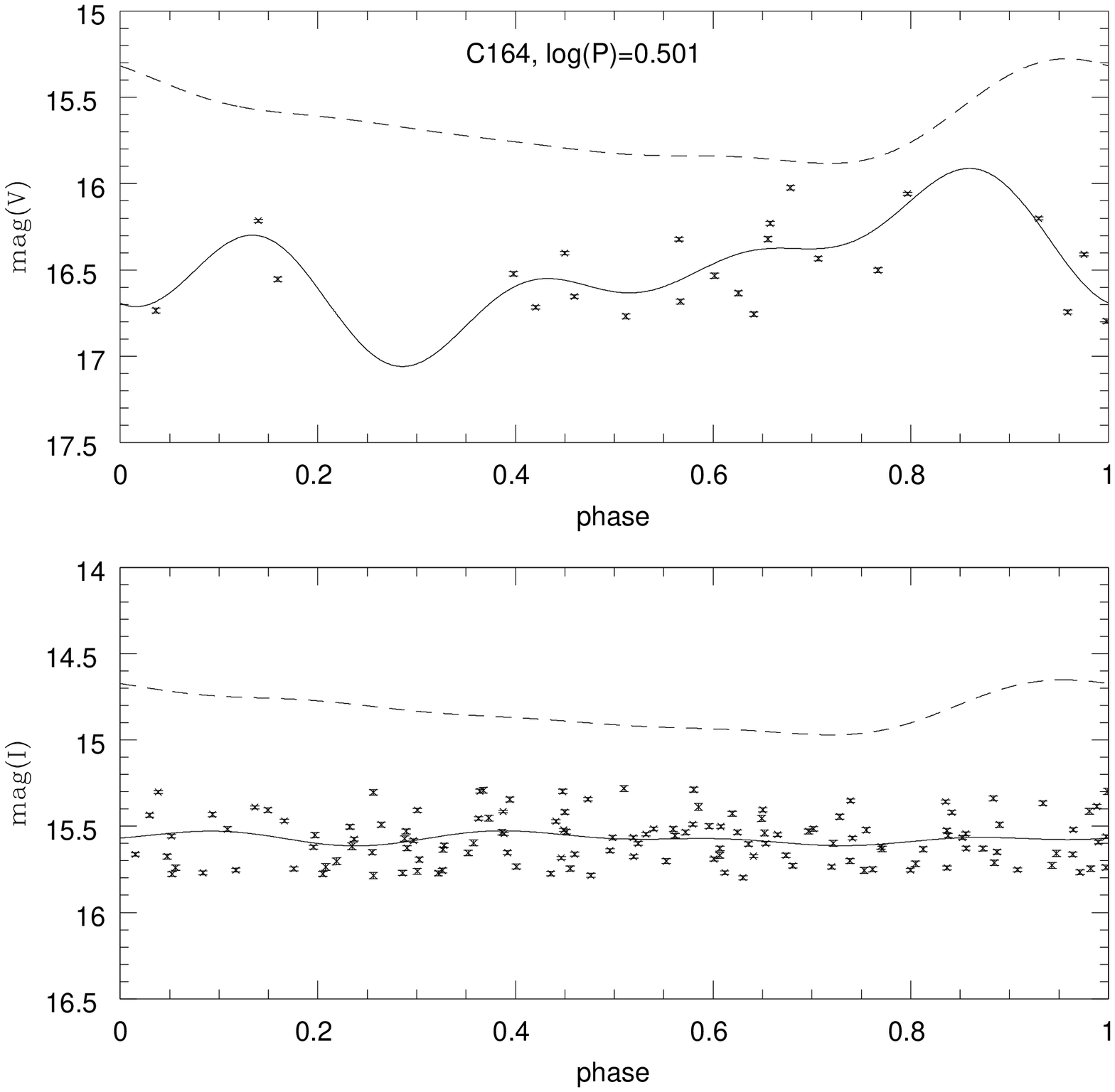}
	\epsfxsize=7.5cm \epsfbox{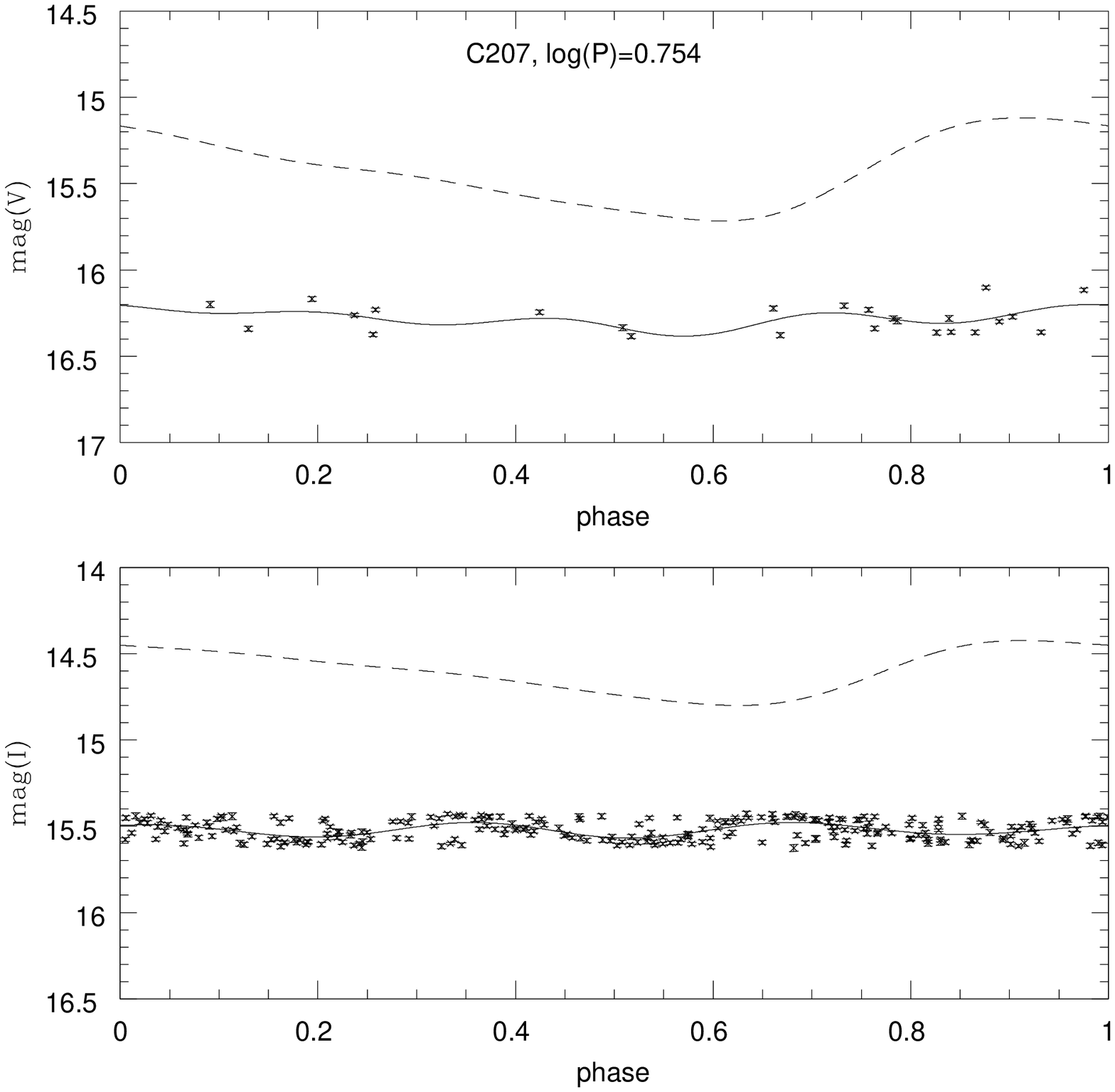}}
	\vspace{0cm}
	\caption{Two examples of unusual light curves in OGLE Cepheids. Upper panels are V band light curve and lower panels are I band light curves. The light curves for Cepheids that have similar periods are shown with dashed lines, for comparison. Original data points are also indicated. \label{fig1}}
	\end{figure*}

%**********************************************************
%      FIGURE 2 HERE
%**********************************************************

	\begin{figure*}
	\vspace{0cm}
	\hbox{\hspace{0.2cm}\epsfxsize=7.5cm \epsfbox{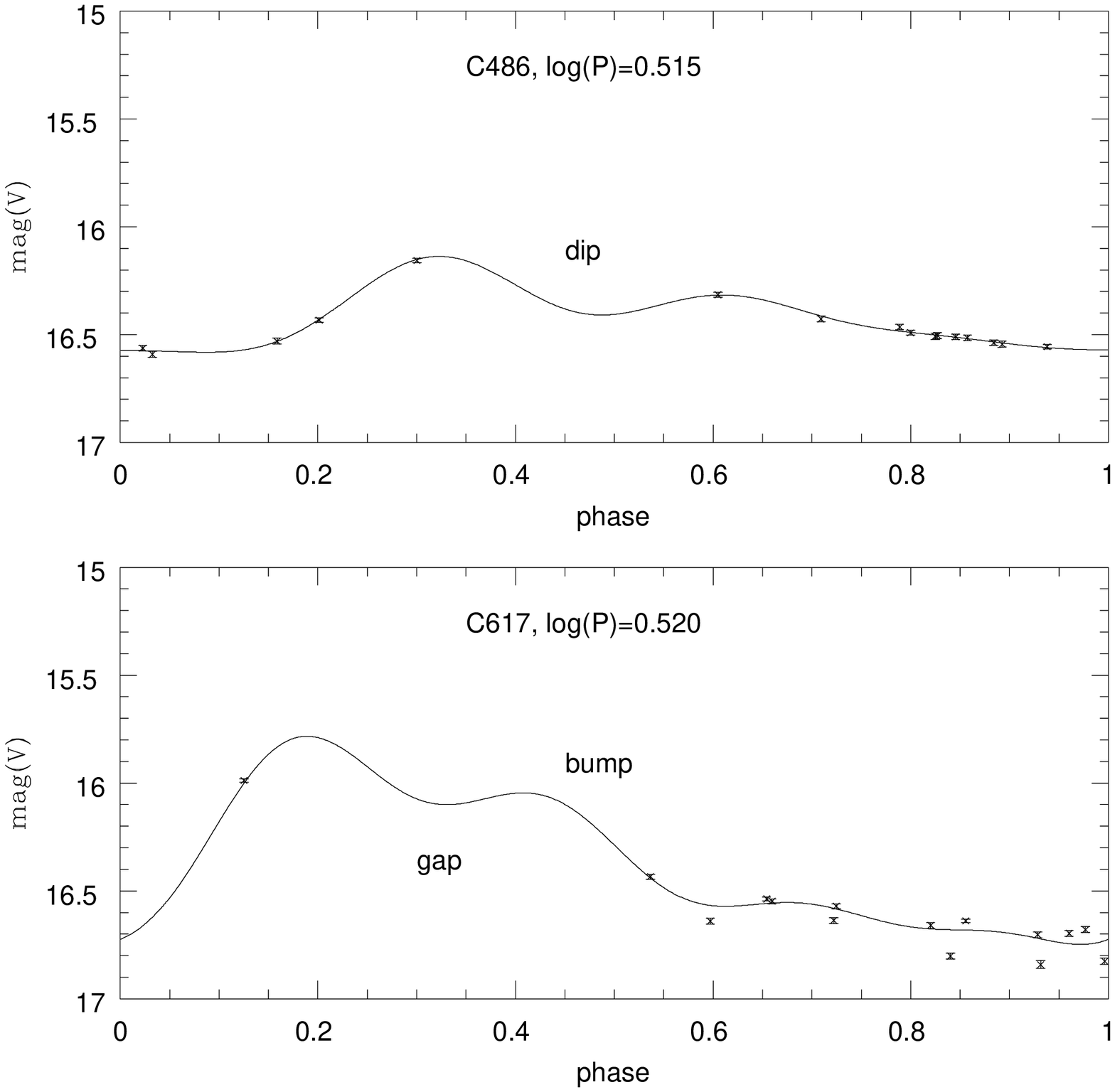}
	\epsfxsize=7.5cm \epsfbox{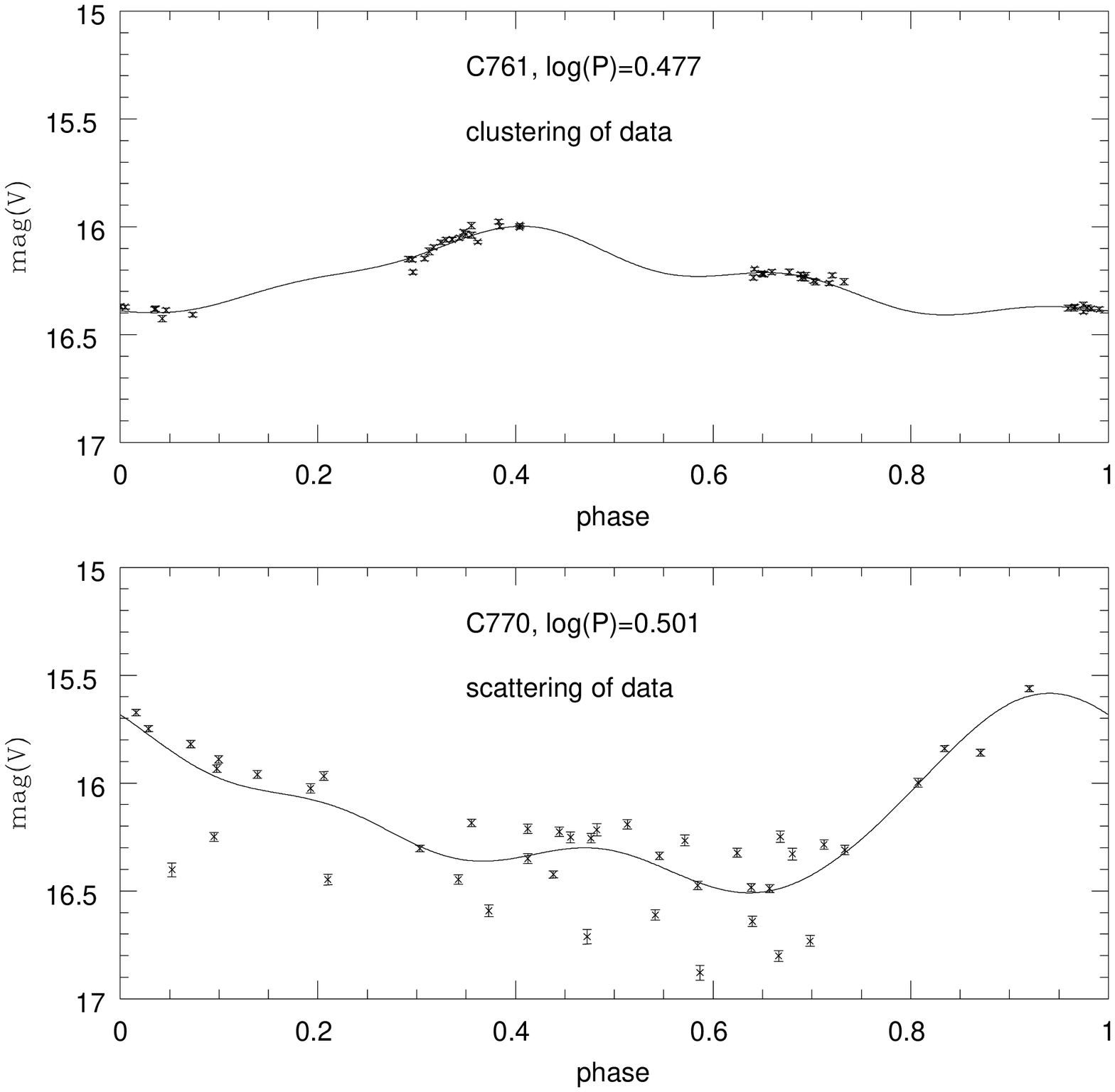}}
	\vspace{0cm}
	\caption{Some examples of poorly constructed V band light curves, due to bad phase coverage or clustering of data points in certain phases. Original data points are also indicated. \label{fig2}}
	\end{figure*}

%**********************************************************
%      FIGURE 3 HERE
%**********************************************************

	\begin{figure*}
	\vspace{0cm}
	\hbox{\hspace{0.2cm}\epsfxsize=7.5cm \epsfbox{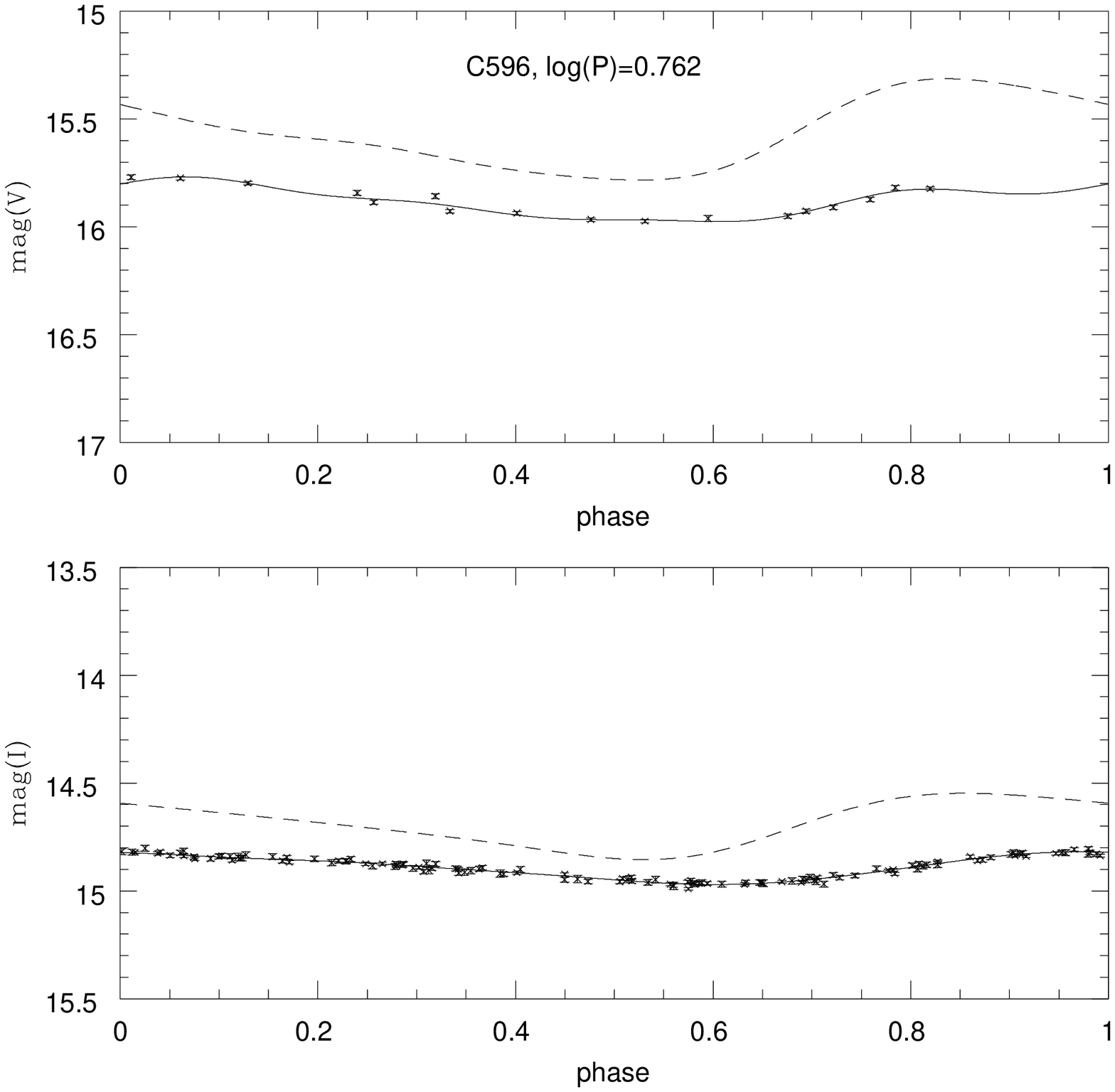}
	\epsfxsize=7.5cm \epsfbox{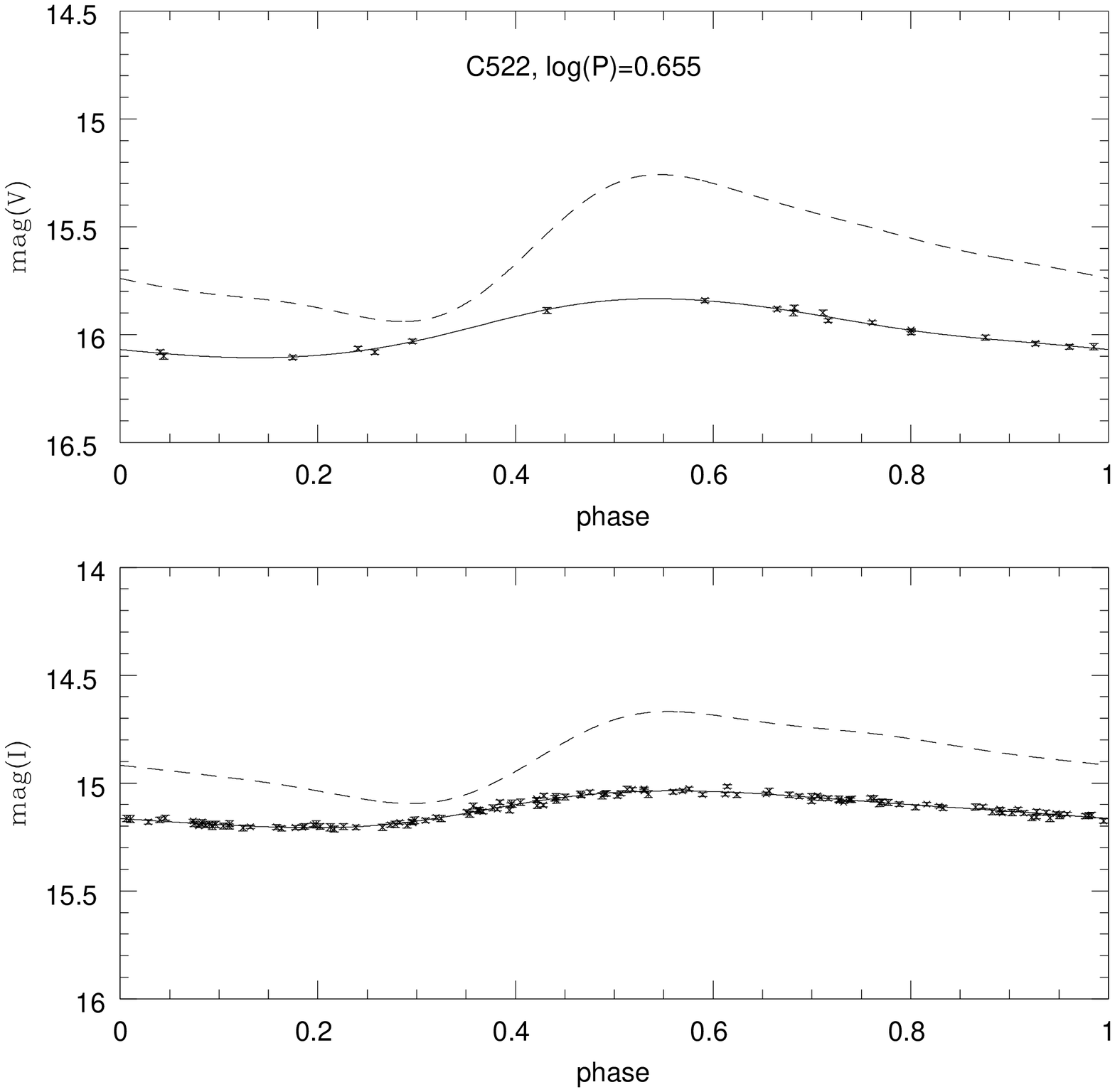}}
	\vspace{0cm}
	\caption{Some examples of Cepheids that have unusually small amplitudes. The light curves for the Cepheids that have similar periods are shown in dashed lines, for comparison. Original data points are also indicated. \label{fig3}}
	\end{figure*}

%**********************************************************
%      FIGURE 4 HERE
%**********************************************************

	\begin{figure*}
	\vspace{0cm}
	\hbox{\hspace{0.2cm}\epsfxsize=7.5cm \epsfbox{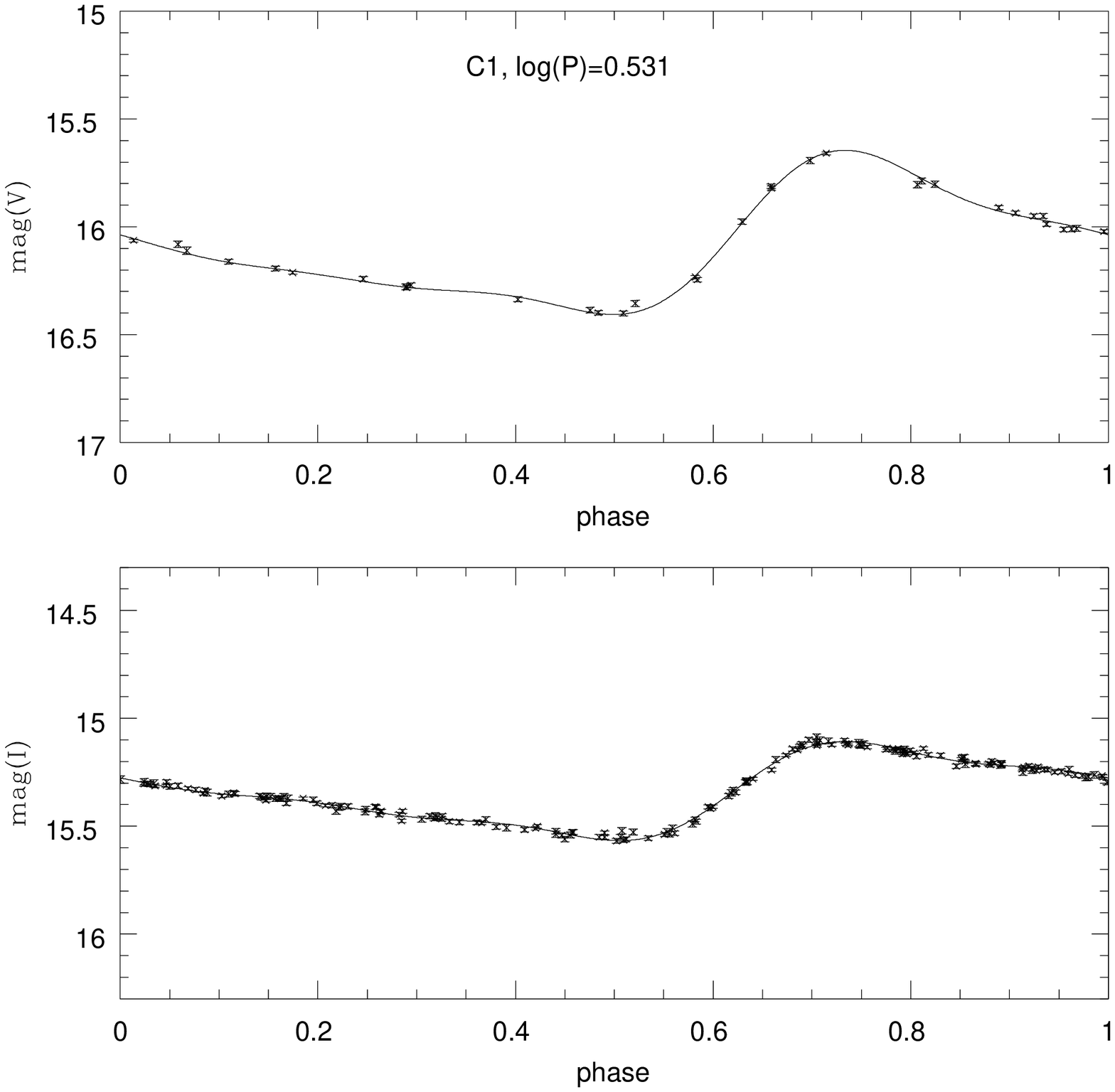}
	\epsfxsize=7.5cm \epsfbox{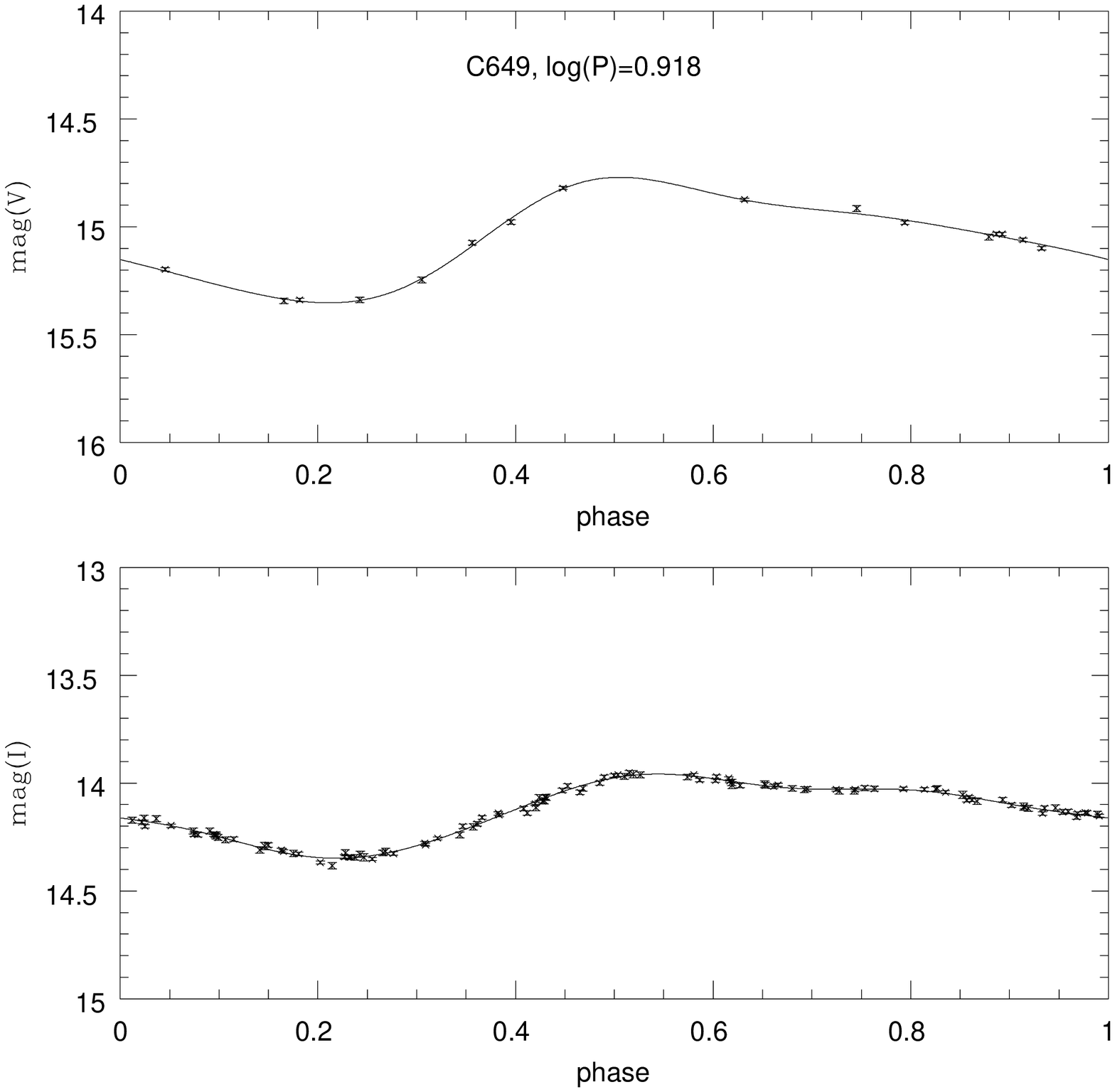}}
	\vspace{0cm}
	\caption{Two examples of well reconstructed light curves. The upper panels are V band light curves and the lower panels are I band light curves.  Original data points are also indicated. \label{fig4}}
\end{figure*}

	We use the same OGLE LMC data as in Udalski et al. (\cite{uda99a}) to derive the LMC PL relations, but without processing the sigma-clipping procedure (Udalski et al. \cite{uda99a}; Willick \& Batra \cite{wb00}; Groenewegen \cite{gro00b}). Instead, we examine in detail the light curves for all Cepheids. The OGLE photometric data were downloaded from the OGLE web-page\footnote{The URL is http://bulge.princeton.edu/$\sim$ogle/}. This consisted of 771 fundamental mode Cepheids (as judged by Udalski et al. \cite{uda99b}). Then, short period Cepheids ($\log(P)<0.4$) were eliminated from the sample, as in the case of Udalski et al. (\cite{uda99a}). The V and I band photometric data for the remaining Cepheids were then fit with the $4^{th}$-order Fourier expansion using simulated annealing techniques (equation (3)) to reconstruct the light curves and obtain the mean magnitudes (Ngeow et al. \cite{nge03}). The light curves for each Cepheid were then visually inspected, and further elimination of data was carried out according to the following criteria:

	\begin{enumerate}
	\item  	In order to have a self-consistent number of Cepheids in both bands, Cepheids without V band data were removed from the sample.
	\item	There are 3 Cepheids that have very unusual light curves, with two of them shown in Figure \ref{fig1}. The nature of these objects or the reason for having the unusual light curves is still unknown. However it is clear that they must be eliminated from the sample.
	\item	Then, we visually examined the V band data and corresponding Fourier expansions for each star to determine which stars had acceptable Fourier expansions. In some cases the V band data are clustered around a phase point resulting in numerical bumps in the Fourier expansion (see Ngeow et al. \cite{nge03} for more details). This is never a problem in the I band because the number of points per light curve is so much greater (more than 120, as compare to V band which has $\sim12$-$\sim50$ data points per light curves). However, the Cepheids with poorly reconstructed V band light curves were eliminated in order to have exactly the same Cepheids in both bands. The number of stars rejected according to this criterion is 34. Some examples of this case are shown in Figure \ref{fig2}. 	
	\item	Finally, some Cepheids appear to have a flat or low amplitude light curves when compared to light curves of stars with similar periods. Figure \ref{fig3} shows a few examples of this case. Since we eliminated the Cepheids with $\log(P)<0.4$, these variables are unlikely to be the RR Lyrae-type variables. Due to the uncertainty of their classification, we eliminated all 43 variables falling into this category. We further discuss this group of Cepheids in Section 4.4.
	\end{enumerate}

%**********************************************************
%      FIGURE 5 HERE
%**********************************************************

	\begin{figure*}
	\vspace{0cm}
	\hbox{\hspace{0.2cm}\epsfxsize=7.5cm \epsfbox{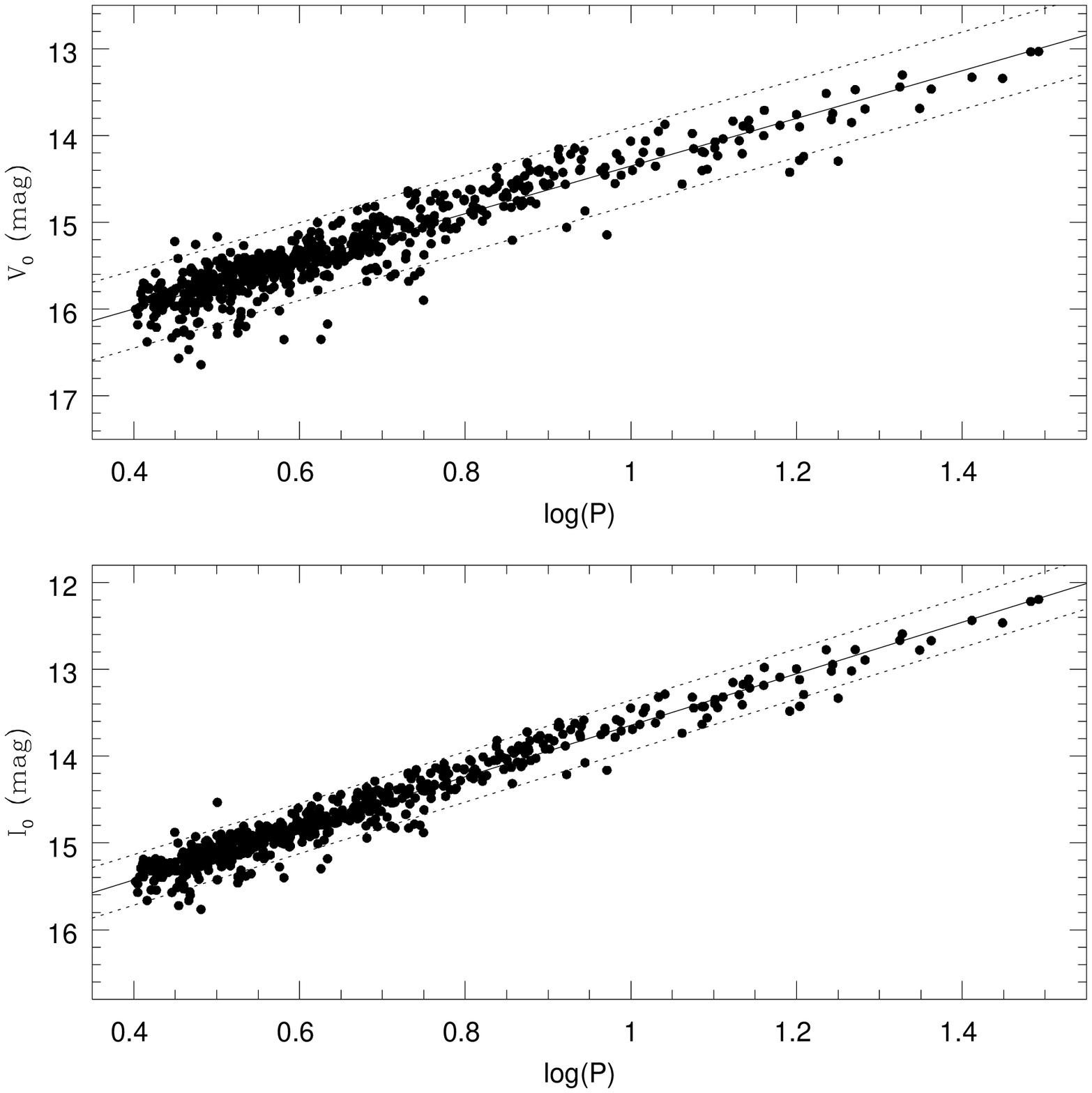}
	\epsfxsize=7.5cm \epsfbox{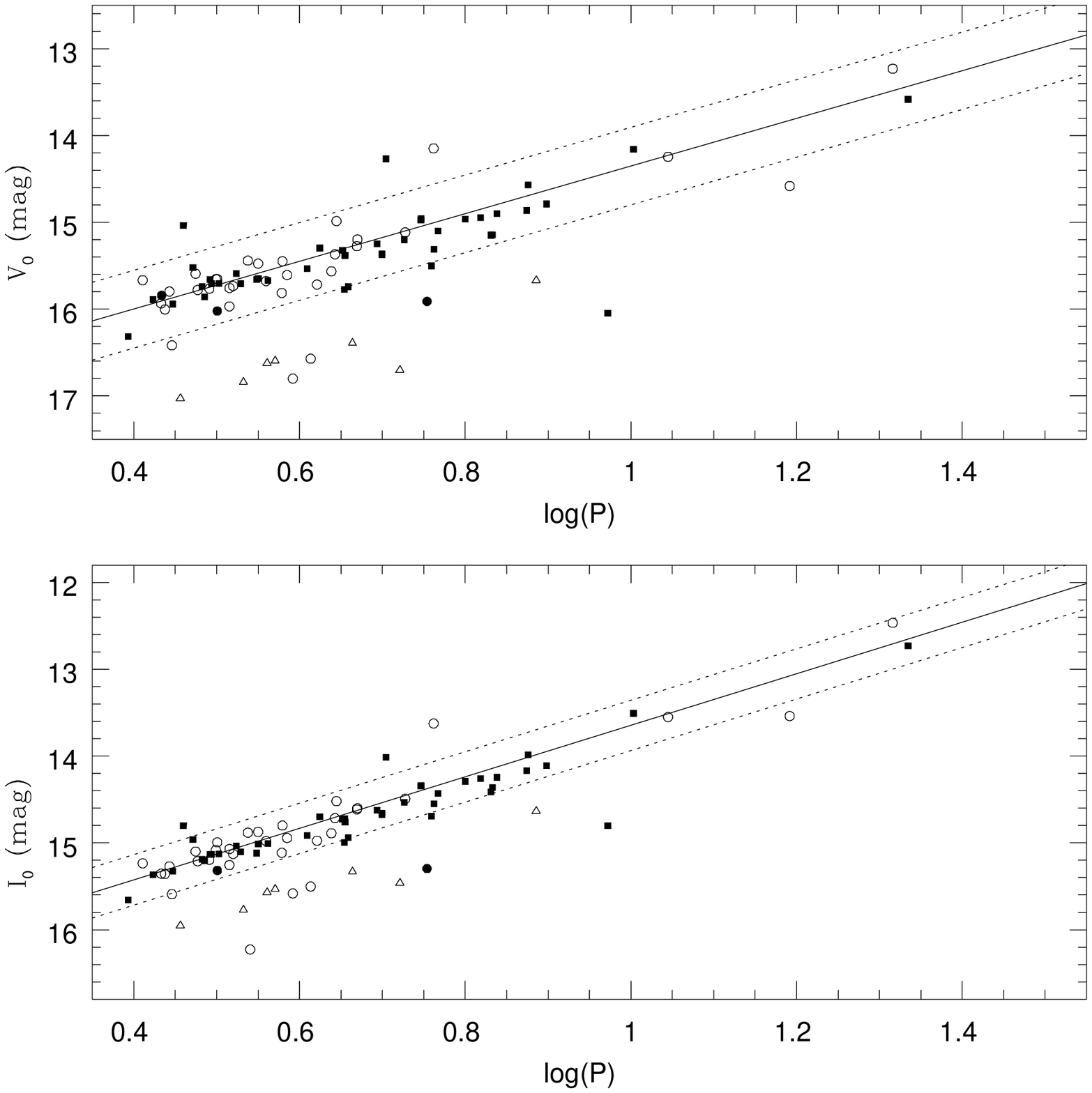}}
	\vspace{0cm}
	\caption{ \textit{Left}: (a) The V and I band PL relations from the final OGLE LMC sample, after correction for reddening. The best fit PL relations are indicated in solid lines. The dashed lines show the $2\sigma$ dispersion, as expected from the width of instability strip. \textit{Right}: (b) The locations of eliminated Cepheids in PL relations. The symbols are: filled circles = 3 Cepheids with unusual light curves (Figure \ref{fig1}); open circles = 34 Cepheids with poorly reconstructed light curves (Figure \ref{fig2}); filled squares = 43 Cepheids with low amplitudes (Figure \ref{fig3}) and open triangles = 7 Cepheids that appear dimmer (by $\sim1.0mag$ in V band) from the ridge line at given periods. \label{fig5}}
	\end{figure*}

	We fit the PL relations to the remaining sample (two examples are shown in Figure \ref{fig4}) with the published extinction values given by OGLE team, assuming $A_V=3.24E(B-V)$ and $A_I=1.96E(B-V)$ (Udalski et al. \cite{uda99b}), to obtain the dereddened PL relations. The extinction map of OGLE LMC Cepheids is derived from observations of red clump stars along 84 lines-of-sight towards the LMC (Udalski et al. \cite{uda99b}). Although there are some criticisms of the OGLE extinction map (e.g., see Feast \cite{fea01}; Beaulieu et al. \cite{bea01}; Fouqu\'{e} \cite{fsg}), we adopt the same extinction map to be consistent with the OGLE team. After fitting the PL relations, there are a few outliers showing up in the plots (not shown) that are more than $\sim4\sigma$ away from the ridge lines at given periods. If we assume a normal distribution, these outliers can be rejected (as in the case of sigma-clipping). However, these outliers are systematically fainter than other Cepheids at similar periods. For example, the V band magnitudes of these outliers are about $1.0mag$ fainter than the mean magnitudes predicted from the ridge lines. Hence, there is a suspicion that these outliers could be the Type II Cepheids (or W Virginis), as they are generally fainter than the (classical) Type I Cepheids.\footnote{The sigma-clipping algorithm will {\it automatically} remove these outliers, but there is no physical reasoning for doing that. In addition, the mean magnitudes for these stars have been corrected for extinction of $A_V\sim 0.5mag$, unless these stars suffer higher extinction than the OGLE extinction map in order to produce the faint magnitudes.} Seven outliers are removed according to this criterion.  

	The resulting PL relations with this final sample are presented in Table \ref{tab1a} and \ref{tab1b}, and the corresponding plots are given in Figure \ref{fig5}(a). In addition, we did not fit the PL relations to long period Cepheids, as in Tammann et al. (\cite{tam01}), although it can be done easily. In Figure \ref{fig5}(b), we also plot out the positions of the eliminated Cepheids in PL plots. As can be seen from the figure, most of the outliers are rejected, as with sigma-clipping. However, some eliminated Cepheids fall along the ridge lines and would not be rejected by the sigma-clipping method. Our philosophy is that it is better to select {\it bona fide} Cepheids, and remove dubious Cepheids that can be eliminated based on physical grounds. Further investigation of the eliminated Cepheids, e.g. the low amplitude Cepheids or the three Cepheids with unusual light curves, are outside the scope of this paper. It is true that in the end there is little difference in our V and I band PL slope relation to that initially published by the OGLE team and used by the KP. Nevertheless, in the age of precision cosmology it is important to use appropriate data. Further our OGLE sample will be important when comparing with pulsation model results. Moreover, with mounting evidence that the PL relation in the LMC is "broken", it is more appropriate to fit two PL relations to the LMC data. However we refit the OGLE data with one relation and included MF91 in our analysis to thoroughly test the effect of using our new way of calculating V and I band means. This is our motivation for studying T02 (the LMC PL relation only for LMC Cepheids with periods longer than 10 days). It must also be pointed out that the calibrating Cepheids in the LMC and Galaxy have periods less than 25 days but many of the extra-galactic Cepheids observed by HST have periods longer than 25 days. This is common to our work and that of KP and STS.

\subsubsection{The Galactic PL Relations}

	Due to high extinction and less accurately known distances to Galactic Cepheids from other independent methods, Galactic PL relations have been considered less favorably in the extra-galactic distance scale problem than LMC PL relations (e.g., Kennicutt et al. \cite{ken95}). However, the idea of using Galactic PL relations has been revived recently (Feast \cite{fea01,fea03}; Tammann \cite{tam03a}; Fouqu\'{e} et al. \cite{fsg}; Thim et al. \cite{thim03}) because not only is the metallicity in many of the HST observed target galaxies closer to Galactic values, but also the current calibration of Galactic PL relations has recently been improved (e.g., see Feast \cite{fea03}; Fouqu\'{e} et al. \cite{fsg}). In addition, because of some possible problems associated with the OGLE LMC PL relations, including the break at 10 days and the dominance of short period Cepheids (see previous subsection), and the possible small ``depth effect'' of LMC Cepheids (Groenewegen \cite{gro00b}; Nikolaev et al. \cite{nik03}), using the Galactic PL relations to calibrate Cepheid distances could be more desirable and can certainly provide an independent check of the LMC based scale which is independent of the LMC distance modulus: this is currently the largest source of systematic uncertainty in the extra-galactic distance scale ($\pm0.1mag$). This does not imply that the systematic uncertainly associated with Galactic PL relations is smaller, but if this is true, the determination of Cepheid distances can be improved.

	We pick three recent Galactic PL relations which have both V and I band PL relations. The first Galactic PL relation is derived from 28 Cepheids using the Barnes-Evans surface brightness technique (Gieren et al. \cite{gie98}, hereafter GFG98). The updated version of GFG98 PL relations has become available just recently. This includes 32 Galactic Cepheids in their sample (Fouqu\'{e} et al. \cite{fsg}, hereafter FSG03). The last Galactic PL relation is adopted from Tammann et al. (\cite{tam03b}), and includes the 28 Cepheids in GFG98 sample with an additional 25 Cepheids from Feast (\cite{fea99}), referred as T03 in Table \ref{tab1a} and \ref{tab1b}. We did not include the Galactic PL relations resulting from the {\it Hipparcos} calibration because these PL relations assumed the V band slopes are the same as in LMC PL relations, and hence calibrate the zero-point only (e.g., see Feast \& Catchpole \cite{fea97}; Lanoix et al. \cite{lan99b}; Groenewegen \& Oudmaijer \cite{gro00a}; Feast \cite{fea03}).  

%*********************************************************************
%  TABLE 4 HERE
%*********************************************************************

	\begin{table*}
	\centering
	\caption{The CTE and metallicity corrections (see text for details).}
	\label{tab2}
	\begin{tabular}{lccccc} \hline
	Galaxy   & $N_{ceph}$ & $[O/H]^{\mathrm a}$ & $\delta_z(LMC)^{\mathrm b}$ & $\delta_z(GAL)^{\mathrm b}$ & $\delta_{CTE}$ \\ \hline \hline
	%\end{tabular}
	%\begin{list}{}{}
	%\item	\centering KP Galaxies
	%\end{list}
	%\begin{tabular}{lccccc} \hline
		&  &  KP Galaxies &  &  &  \\ \hline
	NGC 925	 & 72 & $8.55\pm0.15$ & 0.010 &-0.064 & -0.07  \\
	NGC 1326A& 15 & $8.50\pm0.15$ & 0.0   &-0.074 & -0.07  \\
	NGC 1365 & 47 & $8.96\pm0.20$ & 0.092 & 0.018 & -0.07  \\
	NGC 2090 & 30 & $8.80\pm0.15$ & 0.060 &-0.014 & -0.07  \\
	NGC 2541 & 29 & $8.50\pm0.15$ & 0.0   &-0.074 & -0.07  \\
	NGC 3031 & 17 & $8.75\pm 0.15$ & 0.050 &-0.024 & 0.0$^{\mathrm c}$   \\
	NGC 3198 & 36 & $8.60\pm0.15$ & 0.020 &-0.054 & -0.07  \\
	NGC 3319 & 33 & $8.38\pm0.15$ &-0.024 &-0.098 & -0.07  \\
	NGC 3351 & 48 & $9.24\pm0.20$ & 0.148 & 0.074 & -0.07  \\
	NGC 3621 & 59 & $8.75\pm0.15$ & 0.050 &-0.024 & -0.07  \\
	NGC 4321 & 42 & $9.13\pm0.20$ & 0.126 & 0.052 & -0.07  \\
	NGC 4414 & 8  & $9.20\pm0.15$ & 0.140 & 0.066 & -0.07  \\
	NGC 4535 & 47 & $9.20\pm0.15$ & 0.140 & 0.066 & -0.07  \\
	NGC 4548 & 24 & $9.34\pm0.15$ & 0.168 & 0.094 & -0.07  \\
	NGC 4725 & 15 & $8.92\pm0.15$ & 0.084 & 0.010 & -0.07  \\
	NGC 7331 & 13 & $8.67\pm0.15$ & 0.034 &-0.040 & -0.07  \\ \hline
	%\end{tabular}
	%\begin{list}{}{}
	%\item	\centering STS Galaxies
	%\end{list}
	%\begin{tabular}{lccccc} \hline
		&  &  STS Galaxies &  &  &  \\ \hline
	IC 4182  & 27 & $8.40\pm0.20$ &-0.020 &-0.094 & 0.0$^{\mathrm c}$    \\
	NGC 3627 & 25 & $9.25\pm0.15$ & 0.150 & 0.076 & 0.05   \\
	NGC 3982 & 14 & $(8.9\pm0.4)$ & 0.080 & 0.006 & 0.05   \\	
	NGC 4496A& 45 & $8.77\pm0.15$ & 0.054 &-0.020 & 0.05   \\
	NGC 4527 & 13 & $(8.9\pm0.4)$ & 0.080 & 0.060 & 0.05   \\
	NGC 4536 & 31 & $8.85\pm0.15$ & 0.070 &-0.004 & 0.05   \\
	NGC 4639 & 15 & $9.00\pm0.15$ & 0.100 & 0.026 & 0.05   \\
	NGC 5253 &  5 & $8.15\pm0.15$ &-0.070 &-0.144 & 0.0$^{\mathrm c}$   \\ \hline	
	%\end{tabular}
	%\begin{list}{}{}
	%\item	\centering WM Galaxies
	%\end{list}
	%\begin{tabular}{lccccc} \hline
		&  &  WM Galaxies &  &  &  \\ \hline
	NGC 4258 & 15 & $8.85\pm0.15$ & 0.070 &-0.004 & 0.0$^{\mathrm c}$  \\  \hline
	\end{tabular}
	\begin{list}{}{}
	\item	$^{\mathrm a}$ $[O/H]\equiv 12+\log(O/H)$, adopted from Ferrarese et al. (\cite{fre00}), except for NGC 3982 (Stetson \& Gibson \cite{ste01}) and  NGC 4527 (Gibson \& Stetson \cite{gib01}).
	\item	$^{\mathrm b}$ $\delta_z = (-0.2\pm0.2)\ mag\ dex^{-1} ( [O/H]_{ref} - [O/H])$. For LMC, $[O/H]_{ref}=8.50dex$ and for Galactic, $[O/H]_{ref}=8.87dex$. 
	\item	$^{\mathrm c}$ Since the observations of NGC 3031, IC 4182 and NGC 5253 were done with WFPC (not WFPC2), CTE correction is not included. The reduction of NGC 4258 used Stetson's (1998) calibration (Newman et al. \cite{new01}), no CTE correction is needed.
	\end{list}
	\end{table*}

\subsubsection{LMC PL(Max) Relations}

	Kanbur \& Hendry (\cite{kan96}), on the basis of hydrodynamical pulsation calculations (Simon et al. \cite{sim93}), suggested that PL relations at maximum light may have smaller scatter. However, they found that for a sample of 32 stars in the LMC, the PL relation at maximum light (hereafter PL(Max)) in the LMC had comparable scatter to its counterpart at mean light. Motivated by this, and to use it primarily as a check on our mean light results, we computed the distances to all target galaxies studied in this paper using PL(Max) relations. Maximum light for Cepheids in the target galaxies are estimated using the light curve reconstruction techniques described in Ngeow et al. (\cite{nge03}).

	We use the same OGLE LMC Cepheid sample as in the previous subsection to derive the PL(Max) relations. The estimation of maximum light (or the equivalent minimum magnitude) of a Cepheid from its light curve is more severely influenced by the quality of reconstructed light curves, as compared to its mean light counterpart. This is because bad phase coverage in the data will result in numerical bumps in the reconstructed light curves (Ngeow et al. \cite{nge03}) which can be higher than the maximum light. Therefore, the elimination of poorly reconstructed light curves is more important in this aspect. After the maximum light of each Cepheid in the sample has been estimated from the reconstructed light curves, the PL(Max) relations can be obtained. For this work, we assume that the reddening at maximum light is as same as at mean light even though the period color relations at maximum light are different to those at mean light (Simon et al. \cite{sim93}).

\subsection{CTE and Metallicity Corrections}

	The charge-transfer efficiency (CTE) for WFPC2 is a complicated issue. Simply speaking, it has been found that the performance of the WFPC2 depends on the exposure time, which lead to the ``long-vs-short'' exposure corrections, and other factors (see Stetson \cite{ste98a}; Freedman et al. \cite{kp} for more details). The published photometry data for KP galaxies were based on the Hill et al. (\cite{hil98}) calibration, but the final results presented in Freedman et al. (\cite{kp}) are in the Stetson (\cite{ste98a}) calibration. To convert the calibration from the Hill system to the Stetson system, a correction of $\delta_{CTE}=-0.07\pm0.02mag$ is added to the distance modulus (Mould et al. \cite{mou00}; Freedman et al. \cite{kp}). In contrast, the STS galaxies adopted the Holtzman et al. (\cite{hol95}) calibration, which leads to a correction of $\delta_{CTE}=0.05mag$ (see Saha et al. \cite{sah01a} for example). The calibration used in the WM galaxy is in the Stetson system (Newman et al. \cite{new01}), hence no CTE correction is needed. Note that no CTE correction is added to the three galaxies (IC 4182, NGC 3031 \& NGC 5253) observed with WFPC in the early days of {\it HST}. The CTE corrections for all target galaxies are summarized in the last column of Table \ref{tab2}, and the CTE corrected distance modulus will be represented as $\mu_{0,CTE}$.

	Due to metallicity differences between target galaxies and the LMC, metallicity corrections, as given in equation (2), are commonly applied to $\mu_{0,CTE}$ (Freedman et al. \cite{kp}). The metallicity of the target galaxies, in terms of $[O/H]$, is given in column 3 of Table \ref{tab2}, and the corresponding metallicity corrections with respect to the LMC, $\delta_z(LMC)$, are tabulated in column 4 for the same table. Although the (average) metallicity of the target galaxies is closer to the Galactic value than the LMC (hence the metallicity correction is small), there is no obvious reason for not applying the same metallicity correction when using a Galactic calibrating PL relation. Therefore, we apply the same correction, $\delta_z(GAL)$, to the distance moduli as in the case of using LMC PL relations (thought see Section 4.4). These metallicity corrections are listed in column 5 of Table \ref{tab2} for each target galaxies.

%*********************************************************************
%  TABLE 5 HERE
%*********************************************************************

	\begin{table*}
	\centering
	\caption{The Cepheid distance (with CTE correction) to nearby galaxies with different LMC PL relations.}
	\label{tab3a}
	\begin{tabular}{l|cccc} \hline
	Galaxy  & MF91$^{\mathrm a}$ & U99$^{\mathrm a}$ & Here$^{\mathrm a}$ & TR02$^{\mathrm a}$ \\ \hline \hline
	%\end{tabular}
	%\begin{list}{}{}
	%\item	\centering KP Galaxies
	%\end{list}
	%\begin{tabular}{l|cccc} \hline
	NGC 925	 &  $29.801\pm0.060$ & $29.730\pm0.060$ & $29.718\pm0.060$ & $29.744\pm0.060$ \\
	NGC 1326A&  $31.102\pm0.088$ & $31.007\pm0.088$ & $30.998\pm0.088$ & $31.027\pm0.088$ \\
	NGC 1365 &  $31.315\pm0.052$ & $31.191\pm0.051$ & $31.186\pm0.051$ & $31.218\pm0.052$ \\
	NGC 2090 &  $30.424\pm0.036$ & $30.328\pm0.036$ & $30.319\pm0.036$ & $30.348\pm0.036$ \\
	NGC 2541 &  $30.462\pm0.072$ & $30.355\pm0.071$ & $30.348\pm0.071$ & $30.378\pm0.071$ \\
	NGC 3031 &  $27.887\pm0.094$ & $27.800\pm0.093$ & $27.791\pm0.093$ & $27.818\pm0.093$ \\
	NGC 3198 &  $30.837\pm0.060$ & $30.728\pm0.060$ & $30.721\pm0.060$ & $30.751\pm0.060$ \\
	NGC 3319 &  $30.593\pm0.094$ & $30.511\pm0.092$ & $30.501\pm0.092$ & $30.528\pm0.093$ \\
	NGC 3351 &  $29.979\pm0.090$ & $29.914\pm0.090$ & $29.902\pm0.090$ & $29.927\pm0.090$ \\
	NGC 3621 &  $29.259\pm0.059$ & $29.164\pm0.059$ & $29.155\pm0.059$ & $29.184\pm0.059$ \\
	NGC 4321 &  $30.900\pm0.067$ & $30.769\pm0.067$ & $30.764\pm0.067$ & $30.798\pm0.067$ \\
	NGC 4414 &  $31.236\pm0.052$ & $31.104\pm0.043$ & $31.099\pm0.044$ & $31.133\pm0.045$ \\
	NGC 4535 &  $31.001\pm0.054$ & $30.879\pm0.053$ & $30.874\pm0.053$ & $30.906\pm0.053$ \\
	NGC 4548 &  $30.867\pm0.076$ & $30.783\pm0.078$ & $30.773\pm0.078$ & $30.800\pm0.078$ \\
	NGC 4725 &  $30.522\pm0.066$ & $30.399\pm0.067$ & $30.393\pm0.067$ & $30.426\pm0.067$ \\
	NGC 7331 &  $30.891\pm0.083$ & $30.800\pm0.078$ & $30.791\pm0.079$ & $30.819\pm0.079$ \\ \hline
	%\end{tabular}
	%\begin{list}{}{}
	%\item	\centering STS Galaxies
	%\end{list}
	%\begin{tabular}{l|cccc} \hline
	IC 4182  &  $28.365\pm0.128$ & $28.334\pm0.130$ & $28.318\pm0.130$ & $28.339\pm0.129^{\mathrm b}$ \\
	NGC 3627 &  $30.238\pm0.083$ & $30.134\pm0.084$ & $30.127\pm0.084$ & $30.157\pm0.083$ \\
	NGC 3982 &  $31.714\pm0.136$ & $31.599\pm0.134$ & $31.593\pm0.134$ & $31.624\pm0.135$ \\	
	NGC 4496A&  $30.959\pm0.043$ & $30.839\pm0.043$ & $30.833\pm0.043$ & $30.865\pm0.043$ \\
	NGC 4527 &  $30.755\pm0.129$ & $30.635\pm0.127$ & $30.629\pm0.128$ & $30.661\pm0.128$ \\
	NGC 4536 &  $30.961\pm0.102$ & $30.830\pm0.102$ & $30.826\pm0.102$ & $30.859\pm0.102$ \\
	NGC 4639 &  $31.790\pm0.108$ & $31.664\pm0.105$ & $31.659\pm0.106$ & $31.692\pm0.106$ \\
	NGC 5253 &  $27.881\pm0.244$ & $27.873\pm0.246$ & $27.854\pm0.246$ & $27.872\pm0.245^{\mathrm c}$ \\ \hline
	%\end{tabular}
	%\begin{list}{}{}
	%\item	\centering WM Galaxies
	%\end{list}
	%\begin{tabular}{l|cccc} \hline
	NGC 4258 &  $29.435\pm0.058$ & $29.384\pm0.056$ & $29.370\pm0.056$ & $29.393\pm0.057$ \\ \hline
	\end{tabular}
	\begin{list}{}{}
	\item	$^{\mathrm a}$ MF91 = Madore \& Freedman (\cite{mad91}) PL relations; U99 = Udalski et al. (\cite{uda99a}) PL relations; Here = PL relations derived in Section 3.2.1; TR02= Tammann \& Reindl (\cite{tam02}) PL relations for $\log(P)>1.0$. See Section 3.2.1 for details. 
	\item	$^{\mathrm b}$ There are 12 Cepheids in this galaxy with period less than 10 days. If we use the PL$_{<10d}$ relations, as given by Tammann \& Reindl (\cite{tam02}), then $\mu=28.340\pm0.130$, which is identical to the value with PL$_{>10d}$ relations.
	\item	$^{\mathrm c}$ There are 2 Cepheids in this galaxy with period less than 10 days. If we use the PL$_{<10d}$ relations, as given by Tammann \& Reindl (\cite{tam02}), then $\mu=27.876\pm0.246$, which is very close to the value with PL$_{>10d}$ relations.
	\end{list}
	\end{table*}

%*********************************************************************
%  TABLE 6 HERE
%*********************************************************************

	\begin{table*}
	\centering
	\caption{The Cepheid distance (with CTE correction) to nearby galaxies with different Galactic PL relations.}
	\label{tab3b}
	\begin{tabular}{l|ccc} \hline
	Galaxy  & GFG98$^{\mathrm a}$  & FSG03$^{\mathrm a}$  & T03$^{\mathrm a}$  \\ \hline \hline
	%\end{tabular}
	%\begin{list}{}{}
	%\item	\centering KP Galaxies
	%\end{list}
	%\begin{tabular}{l|ccc} \hline
	NGC 925	 & $29.765\pm0.061$ & $29.768\pm0.060$ & $29.833\pm0.061$  \\
	NGC 1326A& $31.093\pm0.090$ & $31.070\pm0.089$ & $31.166\pm0.091$  \\
	NGC 1365 & $31.336\pm0.053$ & $31.283\pm0.052$ & $31.414\pm0.053$  \\
	NGC 2090 & $30.415\pm0.038$ & $30.392\pm0.036$ & $30.488\pm0.039$  \\
	NGC 2541 & $30.464\pm0.075$ & $30.430\pm0.072$ & $30.539\pm0.075$  \\
	NGC 3031 & $27.867\pm0.098$ & $27.854\pm0.095$ & $27.939\pm0.098$  \\
	NGC 3198 & $30.842\pm0.061$ & $30.805\pm0.060$ & $30.917\pm0.061$  \\
	NGC 3319 & $30.569\pm0.096$ & $30.560\pm0.094$ & $30.639\pm0.096$  \\
	NGC 3351 & $29.936\pm0.090$ & $29.946\pm0.090$ & $30.004\pm0.090$  \\
	NGC 3621 & $29.249\pm0.060$ & $29.226\pm0.059$ & $29.322\pm0.061$  \\
	NGC 4321 & $30.929\pm0.067$ & $30.869\pm0.067$ & $31.008\pm0.067$  \\
	NGC 4414 & $31.265\pm0.063$ & $31.204\pm0.052$ & $31.344\pm0.064$  \\
	NGC 4535 & $31.020\pm0.055$ & $30.969\pm0.054$ & $31.097\pm0.055$  \\
	NGC 4548 & $30.846\pm0.078$ & $30.834\pm0.078$ & $30.917\pm0.078$  \\
	NGC 4725 & $30.541\pm0.067$ & $30.490\pm0.066$ & $30.618\pm0.067$  \\
	NGC 7331 & $30.875\pm0.091$ & $30.858\pm0.083$ & $30.947\pm0.092$  \\ \hline
	%\end{tabular}
	%\begin{list}{}{}
	%\item	\centering STS Galaxies
	%\end{list}
	%\begin{tabular}{l|ccc} \hline
	IC 4182  & $28.285\pm0.129$ & $28.330\pm0.128$ & $28.346\pm0.129$  \\
	NGC 3627 & $30.237\pm0.083$ & $30.206\pm0.083$ & $30.311\pm0.083$  \\
	NGC 3982 & $31.726\pm0.139$ & $31.682\pm0.136$ & $31.802\pm0.139$  \\	
	NGC 4496A& $30.974\pm0.043$ & $30.926\pm0.043$ & $31.051\pm0.044$  \\
	NGC 4527 & $30.771\pm0.132$ & $30.723\pm0.129$ & $30.848\pm0.132$  \\
	NGC 4536 & $30.989\pm0.102$ & $30.929\pm0.102$ & $31.068\pm0.102$  \\
	NGC 4639 & $31.812\pm0.112$ & $31.758\pm0.109$ & $31.890\pm0.113$  \\
	NGC 5253 & $27.776\pm0.243$ & $27.846\pm0.243$ & $27.834\pm0.242$  \\  \hline
	%\end{tabular}
	%\begin{list}{}{}
	%\item	\centering WM Galaxies
	%\end{list}
	%\begin{tabular}{l|ccc} \hline
	NGC 4258 & $29.377\pm0.062$ & $29.401\pm0.058$ & $29.442\pm0.062$  \\  \hline
	\end{tabular}
	\begin{list}{}{}
	\item	$^{\mathrm a}$ GFG98 = Gieren et al. (\cite{gie98}) PL relations; FSG03 = Fouqu\'{e} et al. (\cite{fsg}) PL relations; T03 = Tammann et al. (\cite{tam03b}) PL relations. See Section 3.2.2 for details.
	\end{list}
	\end{table*}

%***********************************************************
%   SECTION 4: RESULTS 
%***********************************************************

\section{Results}
 
	We use the {\it same} Cepheids in each target galaxy that were used by the original KP or STS study after various selection criteria (including the period-cut, color-cut and the quality of the Cepheids) had been applied. The periods of the Cepheids in final samples were taken from the corresponding papers. The mean magnitudes were obtained with the Fourier techniques described in Section 3.1. Then the reddening-corrected distance modulus to individual Cepheids was calculated from equation (1) by using either the LMC or Galactic PL relations. The unweighted mean of the distance moduli to individual Cepheids was taken to be the final distance modulus to the target galaxy. The results of the distance moduli are presented in Table \ref{tab3a} and \ref{tab3b} when using both the LMC and Galactic PL relations respectively. The results obtained from LMC PL(Max) relations are given in Table \ref{tab3c}. Note that the distance moduli in these tables have been corrected for CTE (column 6 in Table \ref{tab2}) but not for metallicity effects, and all the errors in the tables are random (statistical) errors only. 

\subsection{Comparisons of the Mean Magnitudes} 

%**********************************************************
%      FIGURE 6 HERE
%**********************************************************

	\begin{figure*}
	\resizebox{12cm}{!}{\includegraphics{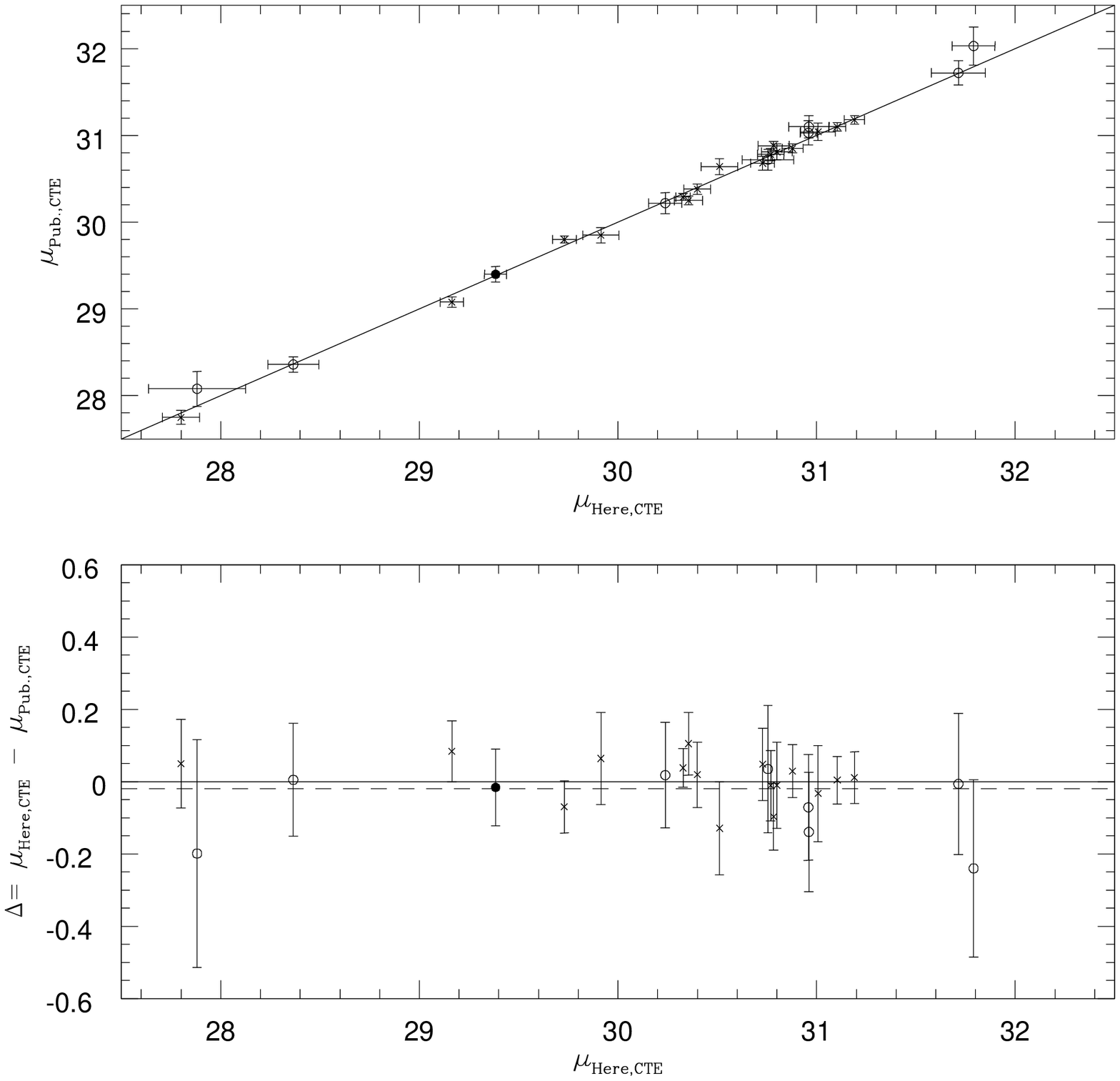}}
	\hfill
	\parbox[b]{55mm}{
	\caption{{\it Upper panel:} Comparison of the distance moduli (after corrected for CTE) obtained by using the Fourier techniques to estimate the mean magnitudes with the published values. The line represents the case of $\mu_{Pub,CTE}=\mu_{Here,CTE}$. {\it Lower panel:} The plot of the difference, $\Delta=\mu_{Here,CTE}-\mu_{Pub,CTE}$, vs. the distance moduli. The dashed line is the average difference of $-0.02$. The symbols are: crosses = 16 KP galaxies; open circles = 8 STS galaxies; and filled circle = WM galaxy.  \label{fig6}} 
	}
	\end{figure*}

	We compare the difference of mean magnitudes obtained with our method to published values. The results of this comparison are presented in Table \ref{tabmag}, where we give the average difference of mean magnitudes ($\Delta/N$) in both bands for each target galaxy. Overall, the differences for the 16 KP galaxies are: $<\Delta V>=0.039\pm 0.012 mag.$ and $<\Delta I>=0.032\pm 0.009 mag.$; for 8 STS galaxies: $<\Delta V>=0.035\pm 0.010 mag.$ and $<\Delta I>=-0.021\pm 0.008 mag.$; and for all 25 target galaxies: $<\Delta V>=0.036\pm 0.008mag.$ and $<\Delta I>=0.014\pm 0.008mag.$
 	
	We also compare our distance moduli, calculated using our Fourier techniques to estimate the V and I band means, with published distance moduli. The only two PL relations we can use in this comparison are the LMC PL relations from MF91 and U99, for STS and KP+WM galaxies, respectively. The results of these comparisons are presented in Table \ref{tab5}, which shows the average difference between our distance moduli and the published values. In overall, the difference is small among all the target galaxies, with a difference of $-0.020\pm0.017mag.$ (corresponding to $\sim$1\% change in distance). This result implies that our method of calculating means is a viable alternative technique, as can be seen from Figure \ref{fig6}. Ngeow et al. (\cite{nge03}) show that our method has advantages in some situations. The largest and smallest difference comes from NGC 4639 and NGC 4414, with a difference of $-0.240\pm0.245mag.$ and $0.004\pm0.066mag.$, respectively.

%**********************************************************
%      FIGURE 7 HERE
%**********************************************************

	\begin{figure*}
	\vspace{0cm}
	\hbox{\hspace{0.2cm}\epsfxsize=7.5cm \epsfbox{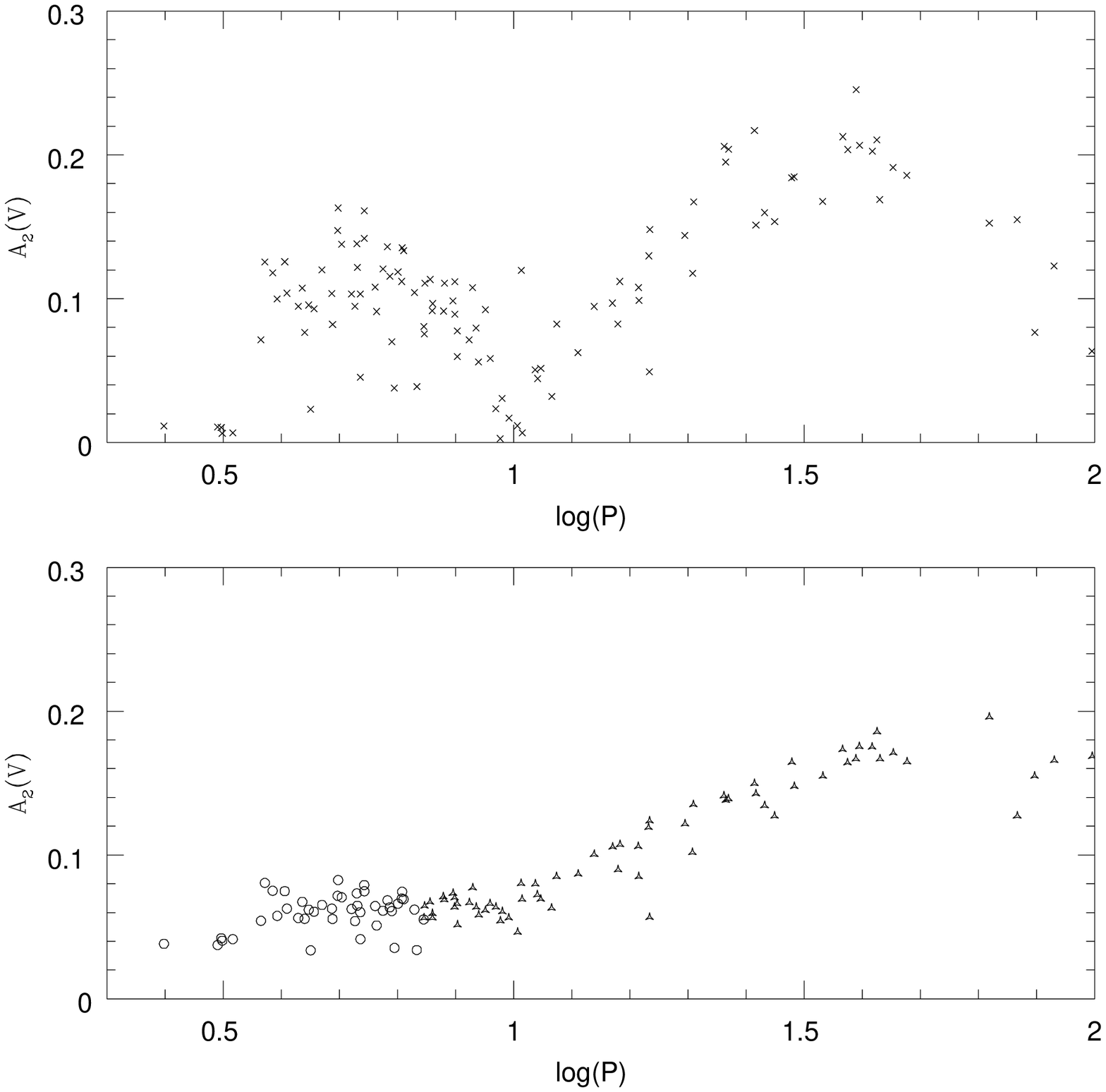}
	\epsfxsize=7.5cm \epsfbox{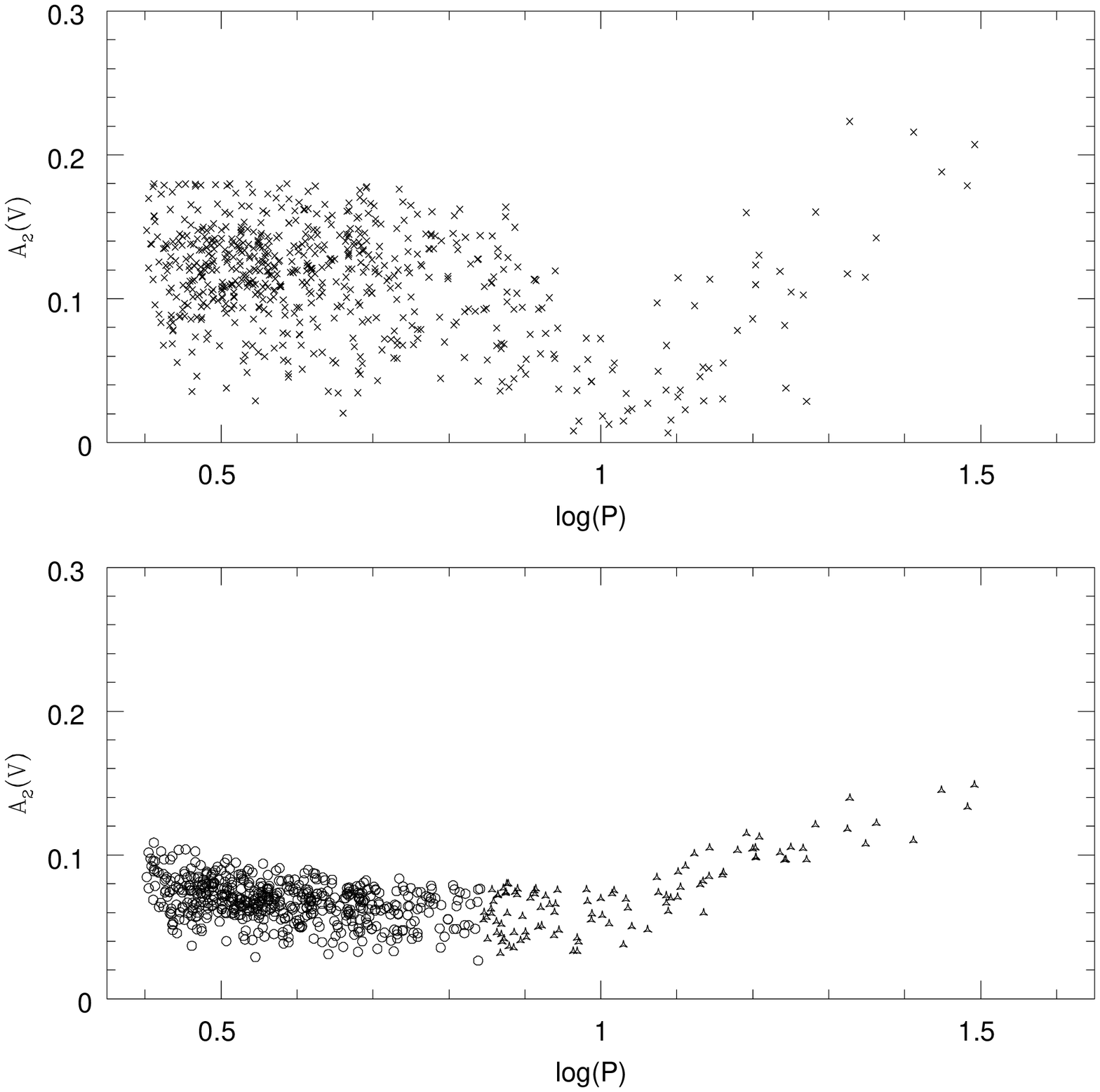}}
	\vspace{0cm}
	\caption{Comparison of $A_2$ from Fourier techniques and template methods. The upper panels show the values of $A_2$ from direct Fourier fit with the method described in Ngeow et al. (\cite{nge03}), and the lower panels show the values of $A_2$ by using the template light curves given by Stetson (\cite{ste96}). Since the template light curves are undefined for $\log(p)<0.85$, Cepheids with period shorter and longer than $\log(p)=0.85$ are represents as open circles and filled triangles, respectively. {\it Left}:(a) The ``calibrating set'' Cepheids. {\it Right}:(b) The OGLE LMC Cepheids. \label{fig9}}
	\end{figure*}

	While STS use phase weighted averages to compute their means, the KP use light curve template techniques (Stetson \cite{ste96}). The difference between this and the method adopted here is described in detail in Ngeow et al. (\cite{nge03}). Both methods fit equations like equation (3) to the observed data. In the case of Stetson (\cite{ste96}), $A_1$ is obtained from the data while $A_k,k>1$ is obtained from the templates showing how the ratio of $A_k/A_1,k>1$ varies with period. In our case all four parameters $A_k,k\ge1$ are obtained from the data with a simulated annealing approach. We do this, firstly, because we want to reconstruct the light curve to model it and, secondly, to capture any light curve shape changes due to metallicity. The top two panels of Figure \ref{fig9} show the second Fourier amplitude coefficient ($A_2$) calculated with this method for very well sampled Galactic (left panel) and LMC (right panel)data. Since this is very well sampled data, it is safe to assume that the progression of $A_2$ with period is a good representation of reality. The bottom panel shows the same Fourier amplitude calculated using the Stetson (\cite{ste96}) method. It can clearly be seen that at all periods the range of $A_2$ at given period is larger than would be predicted by the Stetson (\cite{ste96}) technique. Furthermore, from the well sampled data, we randomly pick 12 points (so that they could be clustered or not as the case may be), add Gaussian errors to these points and then calculate a simulated annealing fit, we recover trends in the top panel of Figure \ref{fig9}. This is clearly displayed in figure 18 of Ngeow et al. (\cite{nge03}).

\subsection{Distance Moduli from Various PL Relations}

%*********************************************************************
%  TABLE 7 HERE
%*********************************************************************

	\begin{table}
	\centering
	\caption{The Cepheid distance to nearby galaxies with PL(Max) relations.}
	\label{tab3c}
	\begin{tabular}{l|c} \hline
	Galaxy  & OGLE LMC PL(Max)$^{\mathrm a}$ \\ \hline \hline
	%\end{tabular}
	%\begin{list}{}{}
	%\item	\centering KP Galaxies
	%\end{list}
	%\begin{tabular}{l|c} \hline
	NGC 925	 & $29.699\pm0.059$ \\
	NGC 1326A& $30.994\pm0.089$ \\
	NGC 1365 & $31.165\pm0.051$ \\
	NGC 2090 & $30.309\pm0.035$ \\
	NGC 2541 & $30.343\pm0.071$ \\
	NGC 3031 & $27.793\pm0.094$ \\
	NGC 3198 & $30.712\pm0.062$ \\
	NGC 3319 & $30.479\pm0.092$ \\
	NGC 3351 & $29.871\pm0.087$ \\
	NGC 3621 & $29.147\pm0.059$ \\
	NGC 4321 & $30.740\pm0.066$ \\
	NGC 4414 & $31.099\pm0.042$ \\
	NGC 4535 & $30.856\pm0.053$ \\
	NGC 4548 & $30.750\pm0.081$ \\
	NGC 4725 & $30.388\pm0.066$ \\
	NGC 7331 & $30.787\pm0.082$ \\ \hline
	%\end{tabular}
	%\begin{list}{}{}
	%\item	\centering STS Galaxies
	%\end{list}
	%\begin{tabular}{l|ccc} \hline
	IC 4182  & $28.253\pm0.130$ \\
	NGC 3627 & $30.116\pm0.087$ \\
	NGC 3982 & $31.592\pm0.143$ \\	
	NGC 4496A& $30.840\pm0.044$ \\
	NGC 4527 & $30.616\pm0.128$ \\
	NGC 4536 & $30.843\pm0.102$ \\
	NGC 4639 & $31.659\pm0.112$ \\
	NGC 5253 & $27.773\pm0.226$ \\ \hline
	%\end{tabular}
	%\begin{list}{}{}
	%\item	\centering WM Galaxies
	%\end{list}
	%\begin{tabular}{l|ccc} \hline
	NGC 4258 & $29.352\pm0.057$ \\  \hline
	\end{tabular}
	\begin{list}{}{}
	\item	$^{\mathrm a}$ Corrected for the CTE effects of $\delta_{CTE}$ (taken from last columns of Table \ref{tab2}).   
	\end{list} 
	\end{table}

	Since the distance moduli listed in Table \ref{tab3a} are calculated using different LMC PL relations, they can be compared. A quick comparison in Table \ref{tab3a} indicates that the MF91 PL relation always give the longer distance moduli, and the shortest distance moduli are from the OGLE LMC PL relations derived in Section 3.2.1. The average difference between the distance moduli derived from MF91 and the OGLE LMC (Here), for all target galaxies in Table \ref{tab3a}, is about $0.105\pm0.006mag.$ Comparing this result with the previous case (i.e. Section 4.1) of using the same PL relations but different means magnitudes, the distance modulus is more sensitive to different calibrating PL relations. Also, the distance moduli from three OGLE LMC PL relations are consistent with each other within the statistical errors. 
	
	A similar situation exists when using Galactic PL relations, as shown in Table \ref{tab3b}. The distance moduli from both GFG98 and FSG03 Galactic PL relations are consistent with each other (with a difference of $\sim 0.024\pm0.007mag.$, as there are 26 common Cepheids in both samples), while FSG03 PL relations produce a shorter distance. However, the distance moduli from T03 PL relations are systematically further than the other two, although all three PL relations share some common Galactic Cepheids. Nevertheless, the average difference between the distance moduli from T03 and FSG03 is $\sim 0.097\pm 0.008mag.$, comparable to the difference seen in LMC PL relations.

	Finally, we compare the distance moduli (after CTE corrections) from LMC PL(Max) relations in Table \ref{tab3c} to their mean light counterparts, i.e., column 4 of Table \ref{tab3a}. The distance moduli from PL(Mean) and PL(Max) relations are consistent with each other, although PL(Max) relations generally give a shorter distance. This serves as an important check on our calibration of the extra-galactic distance scale. The closer distance from PL(Max) relations is expected (with the difference of $\sim 0.015\pm 0.004mag.$, compared to the mean light counterparts), because Cepheids will appear closer at maximum light. The errors associated with maximum light are generally comparable to those at mean light. This will be studied in future work. %This fact is discussed extensively in Kanbur et al. (\cite{kan03}). 

\subsection{Comparison to Published Results}

	In order to compare our results with published values, we need to select the representative distance moduli from both LMC or Galactic PL relations. We pick the distance moduli from the derived OGLE LMC PL relations in Section 3.2.1 (i.e. column 4 of Table \ref{tab3a}) and FSG03 PL relations to represent the LMC and the Galactic results, respectively. The CTE-corrected distance moduli, $\mu_{0,CTE}$, are listed in Table \ref{tab4}. However, the metallicity corrections are not required in these comparisons because they are the same for published results and this work. The effect of metallicity corrections are discussed in Section 4.4.

\begin{figure*}
	\vspace{0cm}
	\hbox{\hspace{0.2cm}\epsfxsize=7.5cm \epsfbox{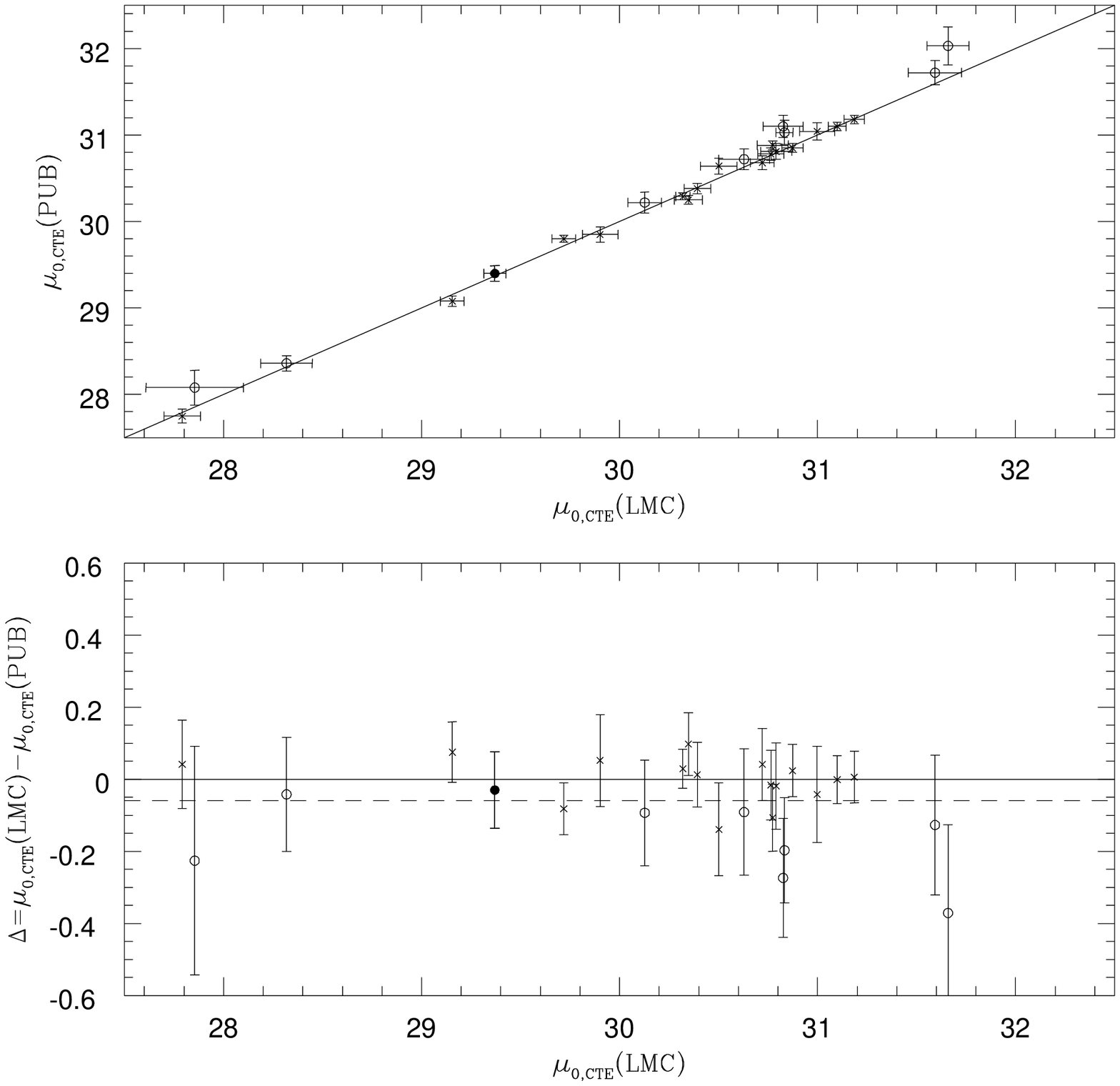}
	\epsfxsize=7.5cm \epsfbox{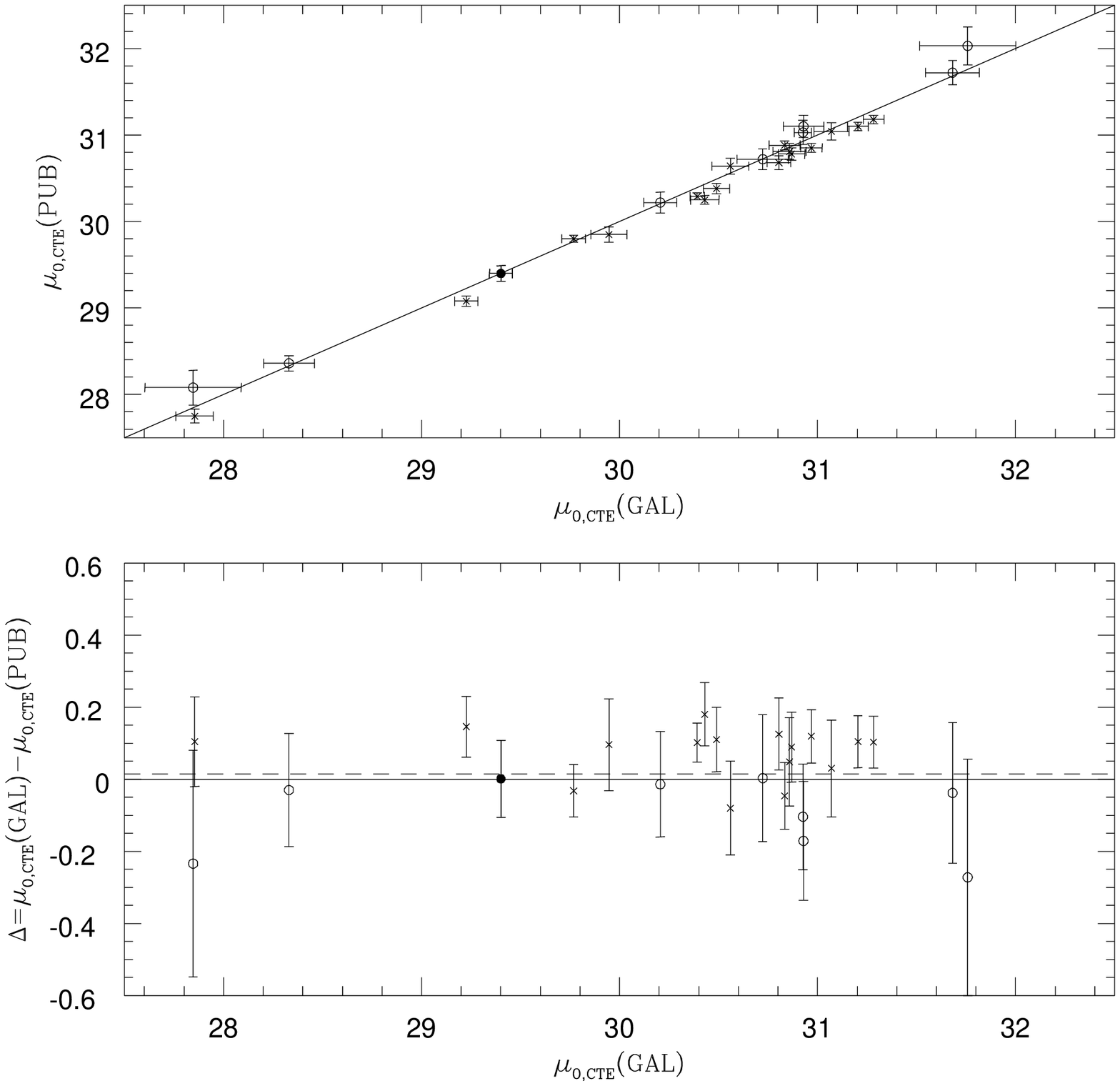}}
	\vspace{0cm}
	\caption{Comparisons with published results. The symbols are same as in Figure \ref{fig6}. {\it Left:(a)} Comparison of our results from the new OGLE LMC PL relations. The dashed line is the average difference of $-0.06$. {\it Right:(b)} Comparison of our results from the Galactic PL relations. The dashed line is the average difference of $0.01$. \label{fig7}}
	\end{figure*}

	The results of the comparisons are plotted in Figure \ref{fig7} and the average difference between our distance moduli and the published values are presented in Table \ref{tab6} for both the LMC and Galactic PL relations. The overall comparisons indicate that our results are consistent with the published results. When using the LMC PL relation, our distance moduli agree well with the KP results, but show a large discrepancy for the STS results ($-0.178 mag.$). A careful comparison of Table \ref{tab3a} with published results shows that this difference arises from two sources: $\sim 0.10 mag.$ is due to the use of MF91 LMC PL relations by STS and $\sim 0.07 mag.$ is due to differences in the V and I band means when calculating the distance moduli (see Section 4.1 and Table \ref{tab5}). However, when using the Galactic PL relations, our results are different by $\sim 0.1mag.$ compared to both KP and STS results. Clearly, the use of Galactic PL relations increases the distance moduli. Surprisingly, the distance modulus of NGC 4258 calculated from LMC PL relations is closer to the water maser distance although the metallicity of this galaxy is nearly identical to Solar.

	Gibson et al. (2000, hereafter G00) used KP techniques to reanalyze the STS galaxies. Since G00 did the photometry from scratch, their list of Cepheids in each target galaxy is in general different from those found by STS. Moreover, they used the Madore \& Freedman (\cite{mad91}) calibrating PL relations (MF91). Gibson \& Stetson (\cite{gib01}) and Stetson \& Gibson (\cite{ste01}) used the same photometry as G00 but used the Udalski et al. (\cite{uda99a}) LMC PL relations (U99) to re-calibrate the distance to all the STS galaxies. They found shorter distance than those published by STS. They suggested that the discrepancy between their work and that of STS is due to the use of U99 LMC PL relations, their photometric analysis of the raw data and the STS analysis to account for possible correlated measurement errors in V and I. While these are important issues, our approach is to use exactly the same photometry as used by both KP and STS groups. Our methods are closer to KP, though we have a different way of estimating the V and I band means. Our results show that, even with exactly the same Cepheids as used by STS, our distance moduli are significantly shorter than STS but significantly longer than those published in Gibson \& Stetson (\cite{gib01}). It would be interesting to perform our analysis on the G00 photometry but this is not available as advertised on the KP web-site.

\subsection{Comparison of Metallicity-Corrected Distance Moduli}

\begin{figure*}
	\vspace{0cm}
	\hbox{\hspace{0.2cm}\epsfxsize=7.5cm \epsfbox{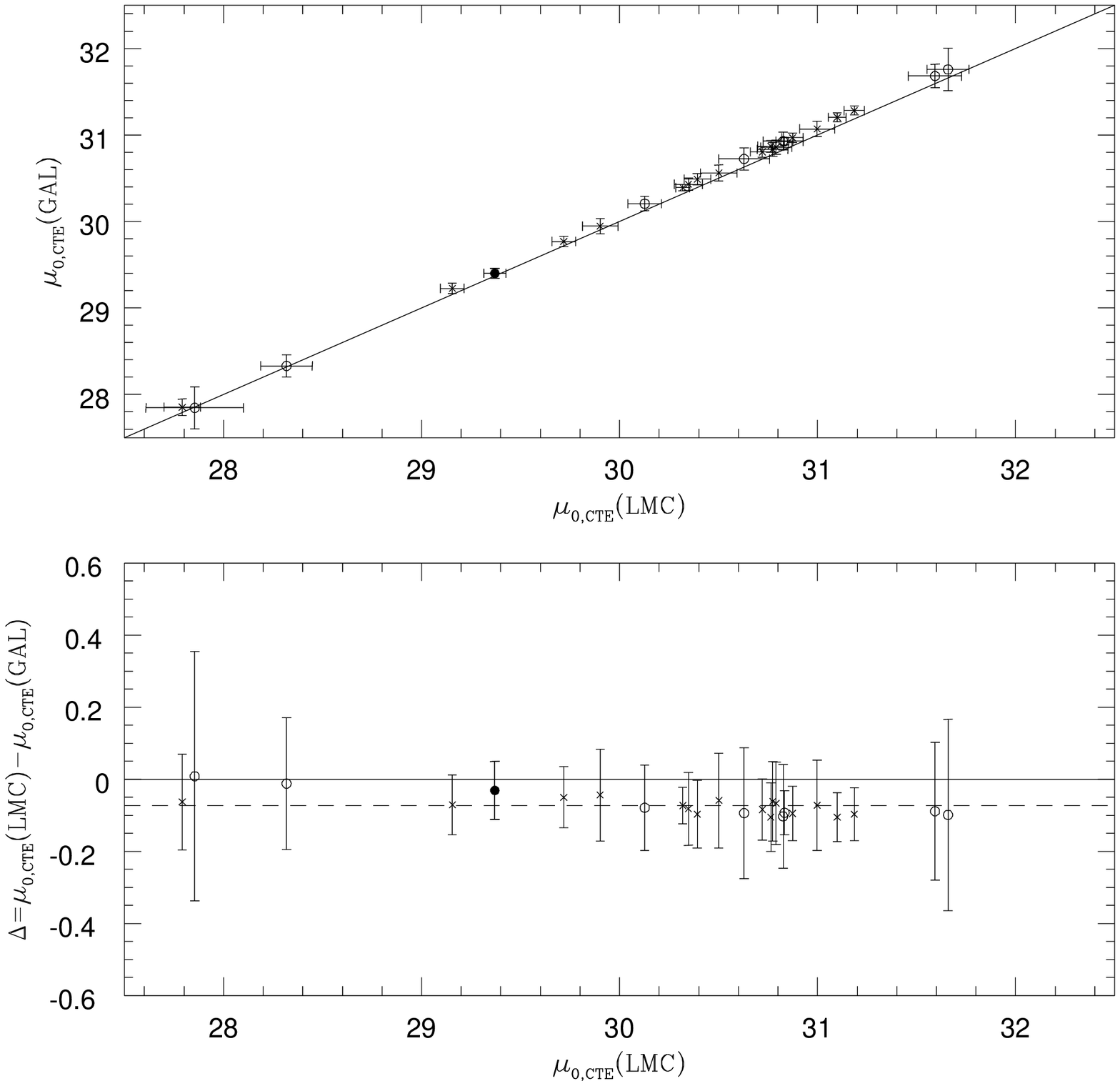}
	\epsfxsize=7.5cm \epsfbox{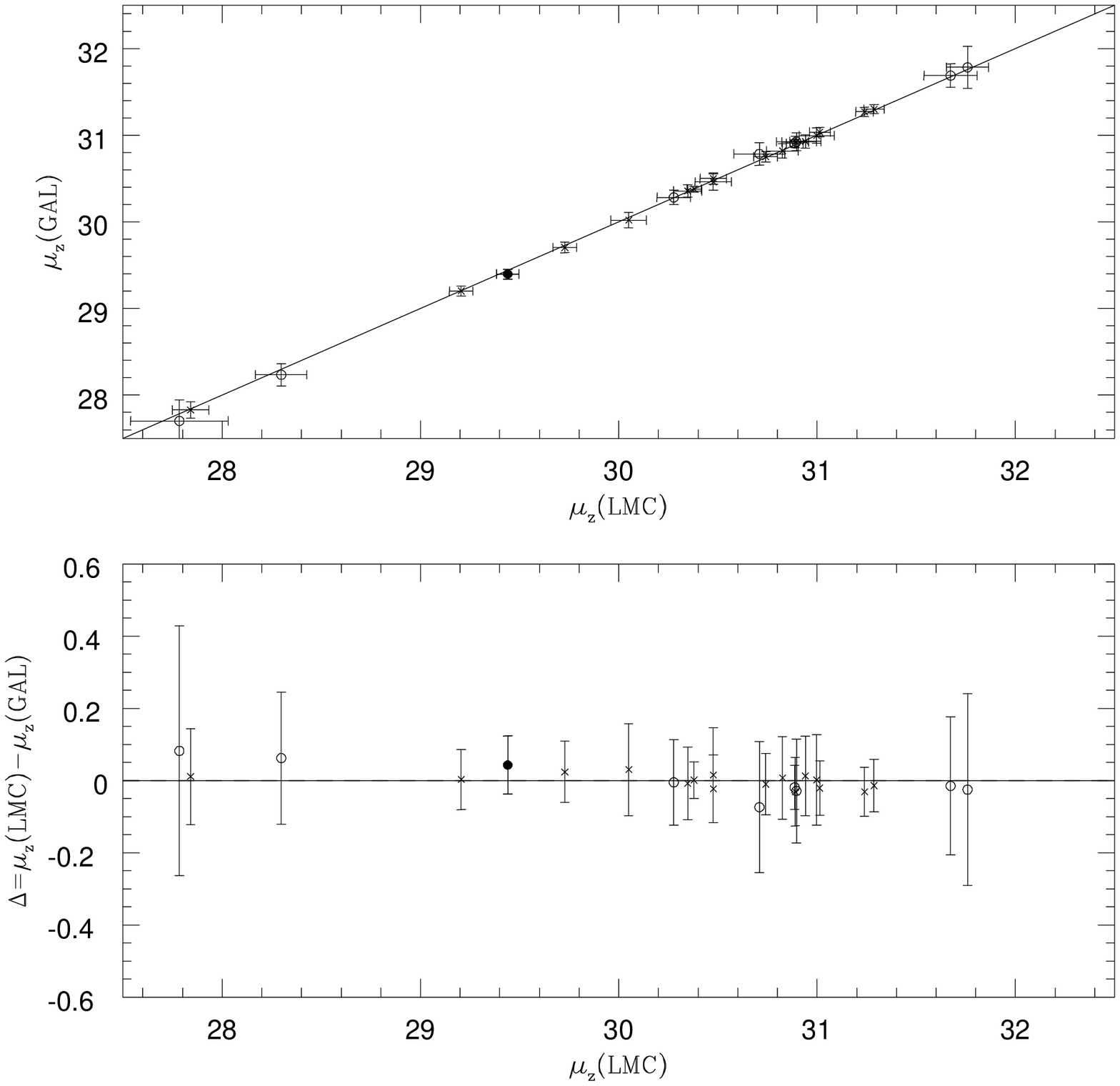}}
	\vspace{0cm}
	\caption{Comparisons of distance moduli from LMC and Galactic PL relations. The symbols are same as in Figure \ref{fig6}. {\it Left:(a)} Uncorrected for metallicity. The dashed line is the average difference of $-0.07$. {\it Right:(b)} Corrected for metallicity. The dashed line is the average difference of $-0.00$. \label{fig8}}
	\end{figure*}

	As well as comparing our results with the published values, we discuss the distance moduli obtained when LMC and Galactic PL relations are used. First, comparing the distance moduli that are uncorrected for metallicity, $\mu_{0,CTE}(LMC)$ and $\mu_{0,CTE}(GAL)$ in Table \ref{tab4}, show that the distance moduli from LMC PL relations are always shorter than their Galactic counterparts (see Figure \ref{fig8}(a)). Summaries of this comparison are given in Table \ref{tab7}. This shows that the LMC PL relations will give a smaller distance modulus by $\sim 0.07\pm0.01mag$ on average (or about 3.5\% in distance) compared to the distance modulus obtained from Galactic PL relations. The negligible random errors suggest that the difference of $0.07mag.$ might be significant, although it is small. However, when the metallicity corrected distance moduli from the two sets of PL relations ($\mu_{z}(LMC)$ and $\mu_{z}(GAL)$ in Table \ref{tab4}) are compared, the difference between them falls close to zero. The results are listed in Table \ref{tab7} and shown in Figure \ref{fig8}(b). 

	Another approach is to use the Galaxy and LMC (P$>10d$) as calibrating PL relations for metal rich and metal poor galaxies respectively. Thus if the seven SN calibrators of STS were forced on an LMC PL relation (U99), one would obtain an average distance decrease of $0.17mag$. However, if we force the five galaxies in their sample, which have on average the same metallicity as the Galaxy, onto the steep relation given in T03, the published STS distances are recovered for these galaxies to within $0.01mag$. Applying the same procedure to the 10 KP galaxies which are metal rich increases their distance moduli by about $0.16mag.$ on average, whilst using TR02 for the remaining metal poor galaxies increases their distance moduli by $0.02mag$.
 
	It has been pointed out by an anonymous referee that it may not be appropriate to eliminate those stars with a low pulsational amplitudes shown in Figure \ref{fig3} as this may bias the LMC PL relation. Such low amplitude Cepheids do exist and are seen in external galaxies. If we include this group of stars in deriving the LMC PL relations, the new PL relations are: $M_V=-2.713(\pm0.044)\log(P)-1.418(\pm0.030)$ and $M_I=-2.943(\pm0.029)\log(P)-1.899(\pm0.020)$. The distance moduli derived from these new LMC PL relations is $\sim0.01mag.$ shorter when compared to the distance moduli derived from LMC PL relations given in Section 3.2.1. When comparing the distance moduli from these new LMC PL relations to those derived using Galactic PL relations, the average difference for the 25 target galaxies considered here is $-0.052\pm0.006mag.$ and $-0.082\pm0.006mag.$ with and without the metallicity corrections respectively. 

        These low amplitude stars are not first overtone pulsators since their light curve Fourier parameters fall in the region occupied by fundamentals. They lie in the instability strip and do not have unusual colors for their period. Thus their luminosities, masses and temperatures are similar to other Cepheids of similar period. So why do they have such a low amplitude? It could be that they have a slightly different composition or are just entering or leaving the fundamental mode instability strip (Buchler \& Koll\'{a}th \cite{buc02}). In this case it is our contention that they should be excluded from the sample since they are Cepheids undergoing a transition.

	The use of Galactic PL relations to calibrate the Cepheid distance scale has been tried before by Paturel et al. (2002a) and Paturel et al. (2002b). The first paper, Paturel et al. (\cite{pat02a}), used the GFG98 sample and the method of ``sosie'' (Paturel \cite{pat84}) to determine Cepheid distances without assuming the PL relations. The second paper, Paturel et al. (\cite{pat02b}), applied the {\it Hipparcos} calibrated PL relations (Feast \& Catchpole \cite{fea97}) to find Cepheid distances. The mean offsets of their results and the published KP results are $0.161\pm0.029mag.$ and $0.027\pm0.016mag.$ for their first and second papers, respectively. The latter result is comparable to our work. However, the extra-galactic Cepheids used in both papers are slightly different than either the KP or STS groups, in contrast with our study in this paper, because these authors applied a different method to deal with incompleteness bias. Also, these authors did not apply any metallicity corrections in their papers.

%*********************************************************************
%  TABLE 8 HERE
%*********************************************************************

	\begin{table*}
	\centering
	\caption{Comparisons of the distance moduli to target galaxies after CTE and metallicity corrections. }
	\label{tab4}
	\begin{tabular}{l|cc|cc|cc} \hline
	Galaxy  & $\mu_{0,CTE}(LMC)$ & $\mu_z(LMC)$ & $\mu_{0,CTE}(GAL)$ & $\mu_z(GAL)$ & $\mu_{0,CTE}(PUB)^{\mathrm a}$ & $\mu_z(PUB)^{\mathrm a}$  \\ \hline \hline
	NGC 925	 & $29.718\pm0.060$ & 29.728 & $29.768\pm0.060$ & 29.704 & $29.80\pm0.04$ & 29.81 \\
	NGC 1326A& $30.998\pm0.088$ & 30.998 & $31.070\pm0.089$ & 30.996 & $31.04\pm0.10$ & 31.04 \\
	NGC 1365 & $31.186\pm0.051$ & 31.287 & $31.283\pm0.052$ & 31.301 & $31.18\pm0.05$ & 31.27 \\
	NGC 2090 & $30.319\pm0.036$ & 30.379 & $30.392\pm0.036$ & 30.378 & $30.29\pm0.04$ & 30.35 \\
	NGC 2541 & $30.348\pm0.071$ & 30.348 & $30.430\pm0.072$ & 30.356 & $30.25\pm0.05$ & 30.25 \\
	NGC 3031 & $27.791\pm0.093$ & 27.841 & $27.854\pm0.095$ & 27.830 & $27.75\pm0.08$ & 27.82 \\
	NGC 3198 & $30.721\pm0.060$ & 30.741 & $30.805\pm0.060$ & 30.751 & $30.68\pm0.08$ & 30.70 \\
	NGC 3319 & $30.501\pm0.092$ & 30.477 & $30.560\pm0.094$ & 30.462 & $30.64\pm0.09$ & 30.62 \\
	NGC 3351 & $29.902\pm0.090$ & 30.050 & $29.946\pm0.090$ & 30.020 & $29.85\pm0.09$ & 30.00 \\
	NGC 3621 & $29.155\pm0.059$ & 29.205 & $29.226\pm0.059$ & 29.202 & $29.08\pm0.06$ & 29.11 \\
	NGC 4321 & $30.764\pm0.067$ & 30.890 & $30.869\pm0.067$ & 30.921 & $30.78\pm0.07$ & 30.91 \\
	NGC 4414 & $31.099\pm0.044$ & 31.239 & $31.204\pm0.052$ & 31.270 & $31.10\pm0.05$ & 31.24 \\
	NGC 4535 & $30.874\pm0.053$ & 31.014 & $30.969\pm0.054$ & 31.035 & $30.85\pm0.05$ & 30.99 \\
	NGC 4548 & $30.773\pm0.078$ & 30.941 & $30.834\pm0.078$ & 30.928 & $30.88\pm0.05$ & 31.05 \\
	NGC 4725 & $30.393\pm0.067$ & 30.477 & $30.490\pm0.066$ & 30.500 & $30.38\pm0.06$ & 30.46 \\
	NGC 7331 & $30.791\pm0.079$ & 30.825 & $30.858\pm0.083$ & 30.818 & $30.81\pm0.09$ & 30.84 \\  \hline
	IC 4182  & $28.318\pm0.130$ & 28.298 & $28.330\pm0.128$ & 28.236 & $28.36\pm0.09^{\mathrm b}$ & $\cdots$ \\
	NGC 3627 & $30.127\pm0.084$ & 30.277 & $30.206\pm0.083$ & 30.282 & $30.22\pm0.12$ & $\cdots$ \\
	NGC 3982 & $31.593\pm0.134$ & 31.673 & $31.682\pm0.136$ & 31.688 & $31.72\pm0.14$ & $\cdots$ \\	
	NGC 4496A& $30.833\pm0.043$ & 30.887 & $30.926\pm0.043$ & 30.906 & $31.03\pm0.14$ & $\cdots$ \\
	NGC 4527 & $30.629\pm0.128$ & 30.709 & $30.723\pm0.129$ & 30.783 & $30.72\pm0.12$ & $\cdots$ \\
	NGC 4536 & $30.826\pm0.102$ & 30.896 & $30.929\pm0.102$ & 30.925 & $31.10\pm0.13$ & $\cdots$ \\
	NGC 4639 & $31.659\pm0.106$ & 31.759 & $31.758\pm0.243$ & 31.784 & $32.03\pm0.22$ & $\cdots$ \\
	NGC 5253 & $27.854\pm0.246$ & 27.784 & $27.846\pm0.243$ & 27.702 & $28.08\pm0.20^{\mathrm c}$ & $\cdots$ \\ \hline
	NGC 4258 & $29.370\pm0.056$ & 29.440 & $29.401\pm0.058$ & 29.397 & $29.40\pm0.09$ & 29.47 \\ \hline
	\end{tabular}
	\begin{list}{}{}
	\item	$^{\mathrm a}$ Published distance moduli for KP galaxies are taken from Freedman et al. (\cite{kp}), Table 4. The distance moduli for STS galaxies and WM galaxy are taken from series of STS papers and Newman et al. (\cite{new01}), respectively.
	\item	$^{\mathrm b}$ Saha et al. (\cite{sah94}) assumed $A_V=A_I=0$, hence $\mu=(\mu_V+\mu_I)/2$.
	\item	$^{\mathrm c}$ The paper (Saha et al. \cite{sah95}) did not list out the final $\mu$, hence this value is calculated via equation (2) with $\mu_V$ and $\mu_I$ given in the paper.
	\end{list} 
	\end{table*}

%**********************************************************
%      FIGURE 8 HERE
%**********************************************************

%**********************************************************
%      FIGURE 9 HERE
%**********************************************************

%****************************************************************
%  SECTION 5: CONCLUSION & DISCUSSION 
%****************************************************************

\section{Conclusion and Discussion}

	In this paper we recalibrate the Cepheid distance to about two dozen nearby galaxies, including the KP and STS galaxies, and a water maser galaxy. We use much of the same methodology as the KP team: the same LMC distance, value of $R$ and metallicity correction, and the same (number of) Cepheids along with the published photometries and periods. However our approach is different from the KP team in two aspects: (a) we estimate the mean magnitudes of sparsely sampled {\it HST} data from Fourier techniques (Ngeow et al. \cite{nge03}) by reconstructing the Cepheid light curves; and (b) we use different sets of PL relations, including the new OGLE LMC PL relations, the Galactic PL relations and the LMC PL(Max) relations, in distance determination. Overall, our results are consistent with each other and KP. We find significantly shorter distances to the STS galaxies. The use of the new Fourier techniques to obtain V and I band means does not produce significant deviations from the existing methods. However the derived distance modulus is more sensitive to the calibrating PL relation that is adopted. The most striking result from this study is that the distance moduli derived from using the LMC and Galactic PL relations are indistinguishable after the metallicity corrections are applied. This provides strong support for the size and quantity of the metallicity dependence of the Cepheid PL relations and the non-universality of PL relations (Tammann et al. \cite{tam03b}). 

	Recent work has shown that the PL relation in the Galaxy and LMC have different slopes and moreover that the PL relation in the LMC is "broken" at a period of 10 days. Metallicity affects the mean brightness of a Cepheid and up to now, this has been the standard motivation for deriving and applying simple additive corrections to the PL relation to account for metallicity differences between calibrating and target galaxy. However it may be that metallicity also affects the slope of the PL relation. If this is so, then the use of a simple additive correction as given in equation (2) is not really appropriate because the PL relation may have a different slope in the target and calibrating galaxy. However, the right panels of figure 8 show clearly that it works in the sense that the distance moduli differences obtained when using the Galactic and LMC as calibrating PL relations are driven close to zero. We try to investigate why this occurs in what follows.

	If we want to compare the distance modulus obtained using LMC and Galactic calibrating PL relations (or any pair of PL relations) when using {\it exactly} the same (number of) Cepheids in the target galaxy, including the same periods and mean magnitudes, then the difference in distance modulus can be expressed as:

	\begin{eqnarray}
	\Delta \mu_0=[(R-1)\Delta a_V-R\Delta a_I]\frac{\sum \log(P)_i}{N}
	\\
	\nonumber 
	\ \ \ \ \ \ +(R-1)\Delta b_V - R\Delta b_I,
	\end{eqnarray}	

	where $\Delta a_{(V,I)}$ and $\Delta b_{(V,I)}$ are the differences in slopes and zero-points for the two PL relations, respectively. Then, for these two PL relations, the change in the distance modulus is a simple linear function of the target galaxy period distribution ($<\log(P)>\equiv \sum \log(P)_i/N$, hereafter mean period), under the assumption of constant $R$ (i.e. the universality of Galactic extinction law, see observational verification for this assumption by Macri et al. \cite{mac01}). The error in $\Delta \mu_0$ is the quadrature sum of the $\sigma_{\mu}$ from both PL relations for the target galaxy. If metallicity corrections of $\delta_z$ (from equation (2)) are applied to equation (5), the metallicity of the target galaxies cancels out and leaves the difference between the LMC and Galactic metallicity. Adopting $[O/H]_{LMC}=8.50dex$ and $[O/H]_{GAL}=8.87dex$, along with $\gamma=-0.2\pm0.2mag.\ dex^{-1}$, the difference in metallicity-corrected distance modulus becomes:

	\begin{eqnarray}
	\Delta \mu_z & = &\Delta \mu_0 - 0.2(8.50-8.87) 
	\\
	\nonumber
	 & = & \Delta \mu_0 + 0.074\ (\pm0.074)
	\end{eqnarray}

	It is worth pointing out that equations (5) and (6), though straightforward to derive, have not been presented in the literature before to the best of our knowledge. Equation (6) is of course dependent on the metallicity correction law adopted and the slopes of the calibrating PL relations. We can re-write equation (5) as $\Delta \mu_0=c<\log(P)>+d$, where $\Delta \mu_0=\mu_0(LMC)-\mu_0(GAL)$, $c=(R-1)\Delta a_V-R\Delta a_I$ and $d=(R-1)\Delta b_V - R\Delta b_I$. For the four LMC PL relations (excluding the PL(Max) relations) and the three Galactic PL relations considered in this paper, there are a total of 12 combinations of $(LMC,GAL)$ PL relations. The coefficients of $c$ and $d$ for each combination are listed in column 2 \& 3 of Table \ref{tab11}. In the same table, we also list out the coefficients of $d$ after applying the metallicity corrections (i.e. equation (6)), i.e. $\Delta \mu_z=c<\log(P)>+d_z$ where $d_z=d+0.074$.

%*********************************************************************
%  TABLE 9 HERE
%*********************************************************************

	\begin{table}
	\centering
	\caption{Comparisons of the mean magnitudes with published results. }
	\label{tabmag}
	\begin{tabular}{l|ccc} \hline
	Galaxy  & $N$ & $\Delta V/N$$^{\mathrm a}$ &  $\Delta I/N$$^{\mathrm a}$ \\ \hline \hline
	NGC 925	 & 72 & $-0.046\pm0.004$ & $-0.006\pm0.010$ \\ 
	NGC 1326A& 15 & $0.040\pm0.009$ & $0.007\pm0.006$ \\
	NGC 1365 & 47 & $-0.077\pm0.006$ & $-0.029\pm0.014$ \\
	NGC 2090 & 30 & $0.079\pm0.004$ & $0.064\pm0.007$ \\
	NGC 2541 & 29 & $0.113\pm0.015$ & $0.086\pm0.020$ \\
	NGC 3031 & 16 & $0.065\pm0.022$ & $0.039\pm0.014$ \\
	NGC 3198 & 36 & $0.001\pm0.059$ & $0.020\pm0.063$ \\
	NGC 3319 & 33 & $0.046\pm0.006$ & $-0.001\pm0.012$ \\
	NGC 3351 & 48 & $0.035\pm0.029$ & $0.009\pm0.012$ \\
	NGC 3621 & 59 & $0.062\pm0.029$ & $0.072\pm0.008$ \\
	NGC 4321 & 42 & $0.018\pm0.009$ & $-0.005\pm0.013$ \\
	NGC 4414 & 8  & $0.027\pm0.007$ & $0.077\pm0.081$  \\ 
	NGC 4535 & 47 & $0.066\pm0.008$ & $0.051\pm0.009$ \\
	NGC 4548 & 24 & $0.087\pm0.021$ & $0.023\pm0.012$ \\ 
	NGC 4725 & 15 & $0.024\pm0.011$ & $0.023\pm0.008$  \\
	NGC 7331 & 13 & $0.086\pm0.010$ & $0.077\pm0.031$ \\ \hline  
	IC 4182  & 26 & $0.050\pm0.013$ & $-0.011\pm0.020$ \\  
	NGC 3627 & 25 & $0.045\pm0.008$ & $-0.018\pm0.026$ \\
	NGC 3982 & 14 & $-0.018\pm0.014$ & $-0.030\pm0.014$ \\
	NGC 4496A& 45 & $0.043\pm0.004$ & $-0.016\pm0.019$ \\
	NGC 4527 & 13 & $-0.013\pm0.016$ & $-0.010\pm0.020$  \\
	NGC 4536 & 31 & $0.057\pm0.005$ & $-0.005\pm0.015$ \\ 
	NGC 4639 & 15 & $0.041\pm0.009$ & $-0.007\pm0.017$ \\
	NGC 5253 &  5 & $0.026\pm0.007$ & $-0.071\pm0.045$ \\  \hline	
	NGC 4258 & 15 & $0.049\pm0.018$ & $0.007\pm0.012$ \\ \hline	
	\end{tabular}
	\begin{list}{}{}
	\item	$^{\mathrm b}$ $\Delta V = V_{Here}-V_{Published}$. Same for $\Delta I$.
	\end{list} 
	\end{table}

%*********************************************************************
%  TABLE 10 HERE
%*********************************************************************

	\begin{table}[h]
	%\centering
	\caption{Comparisons of the distance moduli with different mean magnitudes$^{\mathrm a}$.}
	\label{tab5}
	\begin{tabular}{lc} \hline
	Case & $\Delta (mag.)^{\mathrm b}$  \\ \hline \hline
	16 KP + WM galaxies & $ 0.005\pm0.015$ \\
	8 STS galaxies      & $-0.075\pm0.037$  \\
	All 25 galaxies     & $-0.020\pm0.017$  \\ \hline
	\end{tabular}
	\begin{list}{}{}
	\item $^{\mathrm a}$ By using the same PL relations but different mean magnitudes. See Section 4.1 for details.
	\item $^{\mathrm b}$ Mean Difference, $\Delta=<\mu_{0,CTE}(Here)-\mu_{0,CTE}(Pub.)>$, the errors are the standard deviations of the means.
	\end{list}
	\end{table}

%*********************************************************************
%  TABLE 11 HERE
%*********************************************************************

	\begin{table}[h]
	%\centering
	\caption{Comparisons with the published results.}
	\label{tab6}
	\begin{tabular}{lc} \hline
	Case & $\Delta (mag.)^{\mathrm a}$  \\ \hline \hline
		& Using LMC PL relations       \\ \hline
	16 KP galaxies  & $-0.002\pm0.017$  \\
	8 STS galaxies  & $-0.178\pm0.039^{\mathrm b}$  \\
	All 25 galaxies & $-0.059\pm0.023$  \\ \hline 
		& Using Galactic PL relations  \\  \hline
	16 KP galaxies  & $ 0.075\pm0.018$  \\
	8 STS galaxies  & $-0.108\pm0.038$  \\
	All 25 galaxies & $ 0.014\pm0.024$ \\ \hline
	\end{tabular}
	\begin{list}{}{}
	\item $^{\mathrm a}$ Mean Difference, $\Delta=<\mu_{0,CTE}(Here)-\mu_{0,CTE}(Pub.)>$, the errors are the standard deviations of the means.
	\item $^{\mathrm b}$ This difference is mainly due to the different LMC PL relations used.
	See text (Section 4.3) for details. 
	\end{list}
	\end{table}

%*********************************************************************
%  TABLE 12 HERE
%*********************************************************************

	\begin{table}[]
	%\centering
	\caption{Comparisons of metallicity-corrected distance moduli.}
	\label{tab7}
	\begin{tabular}{lc} \hline
	Case & $\Delta (mag.)^{\mathrm a}$   \\ \hline \hline
		& Without metallicity correction       \\ \hline
	16 KP galaxies  & $-0.077\pm0.005$ \\
	8 STS galaxies  & $-0.070\pm0.015$ \\
	All 25 galaxies & $-0.073\pm0.006$ \\  \hline 
		& With metallicity correction    \\ \hline
	16 KP galaxies  & $-0.002\pm0.005$ \\
	8 STS galaxies  & $-0.003\pm0.018$ \\
	All 25 galaxies & $-0.001\pm0.007$ \\ \hline
	\end{tabular}
	\begin{list}{}{}
	\item $^{\mathrm a}$ Mean Difference, $\Delta=<\mu(LMC)-\mu(GAL)>$, the errors are the standard deviations of the means.
	\end{list}
	\end{table}
	
	To see what values of $<\log(P)>$ would produce identical distance moduli from using either the LMC or the Galactic PL relations, we solve for $<\log(P)>$ such that $\Delta \mu_0=0$ or $\Delta \mu_z=0$. The solutions are listed in columns 5 \& 6 in Table \ref{tab11} for the case of $\Delta \mu_0=0$ and $\Delta \mu_z=0$, respectively. From Table \ref{tab11}, the mean period for two distance moduli to be identical is around $<\log(P)>_0\sim 1.1-1.2$, and $<\log(P)>_z\sim 1.3-1.4$, except for the (MF91,GFG98) and (MF91,FSG03) pairs. In Table \ref{tab12}, we list out the observed mean periods for the target galaxies. This shows that most of the extra-galactic Cepheids in our target galaxies have an observed mean period of $\sim 1.4$ (the mean period for 16 KP galaxies is $1.418$; the mean period for 8 STS galaxies is $1.401$; and the mean period for all 25 galaxies is $1.342$). Therefore, without applying the metallicity correction, the Galactic PL relations will be expected to produce longer distances than the LMC PL relations in most of the (LMC,GAL) pair of PL relations. For example, the change of the distance modulus is $\sim 0.10mag.$ when comparing the LMC PL relations (either U99 or the PL relations derived in Section 3.2.1) to the FSG03 Galacic PL relations in individual target galaxies. However, the LMC and Galactic PL relations will produce almost identical distance moduli after a metallicity correction to within $\sim 0.03mag.$, for most of the target galaxies. The only exception is NGC 5253, because the mean period for this galaxy is 1.029, which is much smaller than the required $<\log(P)>$ of $\sim1.4$.

	The T03 Galactic PL relation has a different slope to the GFG03 and FSG03 Galactic PL relations. Thus the median period in Table \ref{tab11} required for a simple additive correction to be sufficient is approximately 1.2. This is slightly outside the range of mean period required to produce similar distance moduli after a simple metallicity correction when both Galactic and LMC calibrating PL relations are used.

	If the result of this paper, i.e. the near identical distance moduli from LMC and Galactic PL relations after metallicity corrections, is true, then this result can be used to constrain the sign for the coefficient of metallicity correction, the $\gamma$ in equation (2). The value of $\gamma=0.2\pm0.2mag\ dex^{-1}$ used in this study is adopted from Freedman et al. (\cite{kp}), which is roughly the midrange value from several empirical studies (see Section 3). Since the difference of distance moduli between LMC and Galactic PL relations is about $-0.07mag.$ without the metallicity correction (Table \ref{tab7}), a $+0.07mag.$ metallicity correction is required to bring the two distance moduli to be identical. The correction of $0.074mag.$ in equation (6) is almost identical to this requirement, hence the value of $\gamma$ should be around $-0.2mag\ dex^{-1}$ and constrains the sign to be negative. In addition, if the Cepheid PL relations do indeed depend on metallicity, the result of this paper suggests that a simple additive metallicity correction as in equation (2) is a good approximation to model the full complexity of the metallicity dependence of the Cepheid PL relation, provided the mean period of Cepheids in the target galaxy are in the appropriate range for the slopes and extinction laws adopted for the calibrating PL relations\footnote{We thank the referee to point out this.}. Some researchers suggest using the LMC PL relations and Galactic PL relations for metal-poor and metal-rich galaxies, respectively, forgetting about metallicity corrections until an solid understanding of this topic is obtained. This naturally begs the question of which calibrating PL relation to use if the metallicity of the target galaxy is in between the LMC and Galaxy. However our result does not, at the moment, provide evidence supporting one Galactic PL relation over another.

	In summary, the above discussion suggests that: (a) metallicity corrections are necessary when using Cepheid PL relations to find distance moduli; (b) as a consequence, the PL relations do depend on metallicity; (c) hence, Cepheids in the LMC and Galaxy obey different PL relations; (d) the sign for the coefficient of metallicity correction ($\gamma$) has to be negative; and (e) both LMC and Galactic PL relations can be used to determine the distance modulus because either one of the PL relations would yield the same distance modulus after the appropriate metallicity correction. However, further study is needed to test these conclusions.	

%*********************************************************************
%  TABLE 13 HERE
%*********************************************************************

	\begin{table*}
	\centering
	\caption{The change of $\Delta \mu$ with different pairs of LMC and Galactic PL relations$^{\mathrm a}$.}
	\label{tab11}
	\begin{tabular}{lccccc} \hline
	(LMC-GAL) pair& $c$    & $d$   & $d_z=d+0.074$ & $<\log(P)>_0$ & $<\log(P)>_z$ \\ \hline \hline
	(MF91-GFG98)  & -0.257 & 0.369 & 0.443 & 1.4343 & 1.7218 \\
	(U99-GFG98)   & -0.497 & 0.608 & 0.682 & 1.2231 & 1.3719 \\
	(Here-GFG98)  & -0.470 & 0.561 & 0.635 & 1.1946 & 1.3521 \\
	(TR02-GFG98)  & -0.439 & 0.548 & 0.622 & 1.2465 & 1.4149 \\ \hline
	(MF91-FSG03)  & -0.006 & 0.041 & 0.115 & 6.8417$^{\mathrm b}$ & 19.175$^{\mathrm b}$ \\
	(U99-FSG03)   & -0.246 & 0.280 & 0.354 & 1.1392 & 1.4399 \\
	(Here-FSG03)  & -0.218 & 0.233 & 0.307 & 1.0673 & 1.4060 \\
	(TR02-FSG03)  & -0.188 & 0.220 & 0.294 & 1.1678 & 1.5614 \\ \hline
	(MF91-T03)    & -0.300 & 0.365 & 0.430 & 1.1859 & 1.4325 \\
	(U99-T03)     & -0.540 & 0.595 & 0.669 & 1.1018 & 1.2388 \\
	(Here-T03)    & -0.513 & 0.548 & 0.622 & 1.0692 & 1.2135 \\
	(TR02-T03)    & -0.482 & 0.534 & 0.608 & 1.1085 & 1.2620 \\ \hline
	\end{tabular}
	\begin{list}{}{}
	\item	$^{\mathrm a}$ $\Delta \mu = c <\log(P)> + d$ (i.e. equation (5)), and $<\log(P)>$ is the mean period such that $\Delta \mu=0$.
	\item	$^{\mathrm b}$ The large values of $<\log(P)>$ are due to the small value of $c$ in column 2. Hence the MF91 LMC PL relations will always produce a larger distance modulus, as compared to distance modulus from FSG03 Galactic PL relations.
	\end{list}
	\end{table*}

%*********************************************************************
%  TABLE 14 HERE
%*********************************************************************

	\begin{table}[h]
	\caption{The mean period in target galaxies.}
	\label{tab12}
	\begin{tabular}{lc} \hline
	Galaxy   & $<\log(P)>$ \\ \hline \hline
	NGC 925	 & 1.2937 \\ 
	NGC 1326A& 1.3962 \\
	NGC 1365 & 1.5152 \\
	NGC 2090 & 1.3987 \\
	NGC 2541 & 1.4423\\
	NGC 3031 & 1.3577 \\
	NGC 3198 & 1.4525 \\
	NGC 3319 & 1.3398 \\
	NGC 3351 & 1.2681\\
	NGC 3621 & 1.3943 \\
	NGC 4321 & 1.5454 \\
	NGC 4414 & 1.5467 \\ 
	NGC 4535 & 1.5050 \\
	NGC 4548 & 1.3467 \\ 
	NGC 4725 & 1.5081 \\
	NGC 7331 & 1.3733 \\ \hline  
	IC 4182  & 1.1244 \\  
A
	NGC 3627 & 1.4294 \\
	NGC 3982 & 1.4774 \\
	NGC 4496A& 1.4945 \\
	NGC 4527 & 1.4971 \\
	NGC 4536 & 1.5426 \\ 
	NGC 4639 & 1.5201 \\
	NGC 5253 & 1.0292 \\  \hline	
	NGC 4258 & 1.2084 \\ \hline
	\end{tabular}
	%\begin{list}{}{}
	%\item	$^{\mathrm a} <\log(P)>=\sum \log(P)/N$. 
	%\end{list}
	\end{table}

%*********************************************************
%  acknowledgments
%********************************************************

\begin{acknowledgements}
     This work has been supported by NASA Grant GO-09155.03. Part of SN work was performed under the auspices of the U.S. Department of Energy, National Nuclear Security Administration by the University of California, Lawrence Livermore National Laboratory under contract W7405-Eng-48. We thank an anonymous referee for many helpful suggestions that made the paper more relevant. Special thanks to D. Leonard for some discussions and G. A. Tammann for letting us to read their unpublished papers and useful discussion. The authors would like to thank R. Ciardullo for providing the reference of Solar metallicity.
\end{acknowledgements}

%****************************************************
%  REFERENCE 
%***************************************************

\end{document}